\documentclass[
  journal=pasa,
  manuscript=research-paper, 
  year=2020,
  volume=37,
]{cup-journal}

\usepackage{microtype,siunitx,booktabs}

\usepackage{graphicx}
\usepackage{grffile}
\usepackage{amsmath}	
\usepackage{amssymb}	
\usepackage{multicol}       
\usepackage{bm}		
\usepackage{pdflscape}	
\usepackage{float}
\usepackage{hyperref} 
\usepackage[T1]{fontenc}
\usepackage{ae,aecompl}
\usepackage{xcolor} 

\title[Discovery of peculiar radio sources using Machine Learning]{Discovery of Peculiar Radio Morphologies with ASKAP using Unsupervised Machine Learning}

\author{Nikhel Gupta$^{1}$} 
\author{Minh Huynh$^{1,2}$}
\author{Ray P. Norris$^{3,4}$} 
\author{X. Rosalind Wang$^{3}$}
\author{Andrew M. Hopkins$^{5,3}$}
\author{Heinz Andernach$^{6}$}
\author{B\"arbel S. Koribalski$^{4,3}$}
\author{Tim J. Galvin$^{7}$}
\email[Nikhel Gupta]{Nikhel.Gupta@csiro.au}
\affiliation{
$^1$ CSIRO Space \& Astronomy, PO Box 1130, Bentley WA 6102, Australia \\
$^2$ International Centre for Radio Astronomy Research (ICRAR), M468, The University of Western Australia, 35 Stirling Highway, Crawley, WA 6009, Australia \\
$^3$ Western Sydney University, Locked Bag 1797, Penrith, NSW 2751, Australia \\
$^4$ CSIRO Space \& Astronomy, P.O. Box 76, Epping, NSW 1710, Australia \\
$^5$ Australian Astronomical Optics, Macquarie University, 105 Delhi Rd, North Ryde, NSW 2113, Australia \\
$^6$ Depto.\ de Astronom\'{i}a, DCNE, Universidad de Guanajuato, Cj\'on.\ de Jalisco s/n, Guanajuato, CP 36023, Mexico \\
$^7$ International Centre for Radio Astronomy Research, Curtin University, Bentley, WA 6102, Australia
}

\received {02 June 2022}
\revised  {04 August 2022}
\accepted {in PASA 30 August 2022}
\published{31 August 2022}


\usepackage{amsfonts}
\usepackage{color}
\usepackage{mathrsfs}
\usepackage{tabularx}
\usepackage{floatrow}

\def \BE{\begin{equation}}
\def \EE{\end{equation}}	
\def \BC{\begin{center}}
\def \EC{\end{center}}
\def \BEA{\begin{eqnarray}}
\def \EEA{\end{eqnarray}}

\def \SIGMA8{\sigma_{8}}

\usepackage[utf8x]{inputenc}
\usepackage{cellspace}
\begin{document}\sloppy\sloppypar\raggedbottom\frenchspacing


\begin{abstract}
We present a set of peculiar radio sources detected using an unsupervised machine learning method.
We use data from the Australian Square Kilometre Array Pathfinder (ASKAP) telescope to train a self-organizing map (SOM).
The radio maps from three ASKAP surveys, Evolutionary Map of Universe pilot survey (EMU-PS), Deep Investigation of Neutral Gas Origins pilot survey (DINGO) and Survey With ASKAP of GAMA-09 + X-ray (SWAG-X), are used to search for the rarest or unknown radio morphologies.
We use an extension of the SOM algorithm that implements rotation and flipping invariance on astronomical sources.
The SOM is trained using the images of all ``complex'' radio sources in the EMU-PS which we define as all sources catalogued as ``multi-component''.
The trained SOM is then used to estimate a similarity score for complex sources in all surveys.
We select 0.5\% of the sources that are most complex according to the similarity metric, and visually examine them to find the rarest radio morphologies.
Among these, we find two new odd radio circle (ORC) candidates and five other peculiar morphologies.
We discuss multiwavelength properties and the optical/infrared counterparts of selected peculiar sources.
In addition, we present examples of conventional radio morphologies including: diffuse emission from galaxy clusters, and resolved, bent-tailed, and FR-I and FR-II type radio galaxies.
We discuss the overdense environment that may be the reason behind the circular shape of ORC candidates.
\end{abstract}



\section{Introduction}
\label{SEC:Intro}
The next generation of large and deep continuum radio surveys will produce catalogues with multi-million radio sources.
This will have both a huge impact on our understanding of the evolution of galaxies and a large potential for new discoveries.
The majority of these surveys will use advanced radio interferometers, including the Australian Square Kilometre Array Pathfinder \citep[ASKAP:][]{johnston07ASKAP,DeBoer09,hotan21}, 
the Murchison Widefield Array \citep[MWA:][]{tingay13,wayth18}, MeerKAT \citep{jonas16}, the Low Frequency
Array \citep[LOFAR:][]{vanharleem13} and the Karl G. Jansky Very Large Array \citep[JVLA:][]{perley11}.
These instruments have already shown their capability to survey hundreds of square degrees of radio sky at unprecedented depths.
To capture the full potential of these surveys comes the need to transform the data analysis and interpretation techniques.

Historically, the greatest scientific discoveries with major telescopes are serendipitous and lie beyond the original goals of the experiment \citep[][]{norris15}.
\cite{ekers09} finds that in the last 60 years, only seven out of 18 major astronomical discoveries were planned.
Currently, existing methods to make unexpected discoveries are primarily powered by human intelligence that are not expected to scale up to the massive data volumes of this decade.
Without redesigning the search efforts, several unknown radio phenomena may take years to be found, or may never be found.

In recent years, machine learning has emerged as a powerful tool to model highly non-linear data.
Depending on the availability of data, machine learning can be performed in a supervised or unsupervised manner.
For supervised learning the model is trained on several examples of input-output pairs.
Such a model trained with truth labels is then used to estimate the output from a given input.
Recently, these machine learning models have shown encouraging results when used to classify the radio source morphologies \citep[e.g.][]{lukic18, alger18,wu19,viera21}.
However, without training labels these models in their current form are useless.
With multi-million radio detections in future surveys where labelling a large training dataset is both expensive and time consuming, making it more pertinent to invest in unsupervised learning techniques.

In the present work, we use a self-organizing map \citep[SOM][]{kohonen82} that does not require truth labels and focuses on the recognition of structure in a dataset.
SOMs have previously been used to classify the radio morphologies \citep[e.g.][]{ralph19,galvin19,galvin20} and very recently to find some of the rarest radio morphologies \citep[][]{mostert21}.
Following these previous studies, we use an implementation of the SOM that is invariant to affine transformations e.g. rotational, flipping and scaling variation of a radio galaxy.
We train a SOM using a catalogue of ``complex'' (defined here as all multi-component) sources in ASKAP's Evolutionary Map of Universe pilot survey \citep[EMU-PS;][]{norris11, norris21}.
The trained SOM is then used to find the most unusual radio sources.
We derive a similarity metric for complex sources in EMU-PS as well as the pilot phase of Deep Investigation of Neutral Gas Origins survey (DINGO\footnote{https://dingo-survey.org/}) and the Survey With ASKAP of GAMA-09 + X-ray \citep[SWAG-X;][]{moss22prep}.
Based on this similarity metric score, we visually inspect sources with the top 0.5\% most complex radio morphologies. 
We present the rarest radio morphologies in the top 0.5\% complex sources.
Among these are the peculiar morphologies with unusual radio structures and, no corresponding diffuse emission in the optical wavelengths.
We briefly discuss some of these peculiar sources in the present paper and note that future work should study them in more detail to understand the unconventional physical mechanisms behind their formation.
In addition, the rest of the top 0.5\% complex sources have conventional radio morphologies with known mechanisms of formation.
We present few examples of these sources as well.

The paper is structured as follows.
In Section~\ref{SEC:Observations}, we describe the ASKAP observations and other multiwavelength datasets we used.
Section~\ref{SEC:method} is dedicated to the methods that include data pre-processing, description of SOMs, details about the network training, and the procedure to select peculiar sources.
In Section~\ref{SEC:results}, we present a multiwavelength view of peculiar radio sources and examples of conventional sources.
In Section~\ref{SEC:Discussion}, we discuss the overdensity of galaxies near the circular radio sources.
We summarise our findings in Section~\ref{SEC:Summary} and provide directions for future work.
Throughout this paper, we assume a flat $\Lambda$CDM cosmology based on \cite[][]{planck18-1} with $H_0=67.5$ and $\Omega_{m}=0.315$.

\section{Observations}
\label{SEC:Observations}
In this section we describe the radio, infra-red and optical observations we used.
\subsection{ASKAP Observations}
\label{SEC:ASKAP}
ASKAP is a radio telescope located at the Murchison Radio-astronomy Observatory (MRO). 
The telescope is equipped with the phased array feed \citep[PAF:][]{hay06} technology that enables high survey speed by virtue of wide instantaneous field of view. 
ASKAP has 36 antennas with a range of baselines. 
Most of these are located within a region of 2.3 km diameter, with the outer six extending the baselines up to 6.4 km \citep{hotan21}.
ASKAP has recently completed the first all-sky Rapid ASKAP Continuum Survey \citep[RACS:][]{McConnell20} covering the entire sky south of Declination $+41^{\circ}$ to a median RMS of about 250 µJy/beam.
This has paved a way for subsequent deeper surveys using ASKAP.

One such survey is the Evolutionary Map of the Universe \citep[EMU;][]{norris11}, which is planned to observe the entire Southern Sky and is expected to produce a catalogue of as many as 40 million sources \footnote{Forecast based on the allocated time for the EMU 5-year survey program (see https://www.atnf.csiro.au/projects/askap/commissioning\_update.html).}.
Proceeding in this direction, the EMU Pilot Survey \citep[EMU-PS:][]{norris21} was completed in late 2019.
The EMU-PS covers 270 deg$^2$ of sky with $301^{\circ}< {\rm RA} < 336^{\circ}$ and $-63^{\circ}< {\rm Dec} < -48^{\circ}$.
It consists of 10 tiles with total integration time of $\sim 10$ hours each, reaching an RMS sensitivity of $25-35~\mu$Jy/beam and a beamwidth of $13^{\prime\prime} \times 11^{\prime\prime}$ FWHM.
The operating frequency of EMU-PS is between 800 and 1088 MHz centred at 944 MHz.
The raw data was processed using the ASKAPsoft pipeline \citep[][]{whiting17,norris21}. 
As the survey data consists of ten overlapping tiles, value-added processing was performed to produce a unified image and source catalogue.
This includes merging of tiles by performing the weighted average of the data in overlapping regions and convolving the unified image to a common restoring beam size of $18^{\prime\prime}$ FWHM to overcome the variations in point spread function (PSF) from beam to beam \citep[][]{norris21}. 
A catalogue of islands and components is then constructed by running the ''$Selavy$" source finder \citep[e.g.][]{whiting12} on the convolved image.
This catalogue contains 220,102 components with 81.3\% simple (or single component) and 18.7\% complex (multiple components) sources.
As the main goal of the present work is to find a way to streamline a search of new peculiar radio sources, we have limited our analysis to the 41,181 components of complex sources in the catalogue.

The second survey used here is the Deep Investigation of Neutral Gas Origins pilot survey (DINGO\footnote{https://dingo-survey.org/}). 
DINGO aims to provide a legacy of deep HI observations out to redshift $z\sim0.4$. 
The key science goals of DINGO are to study the evolution of the cosmic HI density and the evolution of galaxies \citep[][]{meyer09}. 
The central frequency of the survey is 1367 MHz.
In the present work, we use 11 DINGO tiles publicly available from the CSIRO ASKAP Science Data Archive (CASDA\footnote{https://research.csiro.au/casda/}).
Each tile has a total integration time of $\sim 8$ hours except for two tiles with $\sim 6$ hours of integration.
The average beamwidth of the survey is $10^{\prime\prime} \times 6^{\prime\prime}$ FWHM.
Each tile was processed using ASKAPsoft with standard continuum settings.
Seven tiles with Scheduling Block IDs (SBIDs): 10991, 10994, 11000, 11003, 11006, 11010 and 11026 cover the same sky region with $338^{\circ}< {\rm RA} < 346^{\circ}$ and $-36^{\circ}< {\rm Dec} < -29^{\circ}$.
These tiles have RMS sensitivity between 49 and $64~\mu$Jy/beam.
Weighting the individual tiles proportional to $1/\rm RMS^2$ we generate an averaged map from these tiles with a final RMS sensitivity near $21~\mu$Jy/beam.
In the same way, tiles with SBIDs 14109 and 14136 covering the area of $217^{\circ}< {\rm RA} < 223^{\circ}$ and $-3^{\circ}< {\rm Dec} < +4^{\circ}$ are also combined to get a second averaged map with final RMS noise of $40~\mu$Jy/beam.
A third averaged map is generated combining SBIDs 14055 and 14082 covering the area of $211^{\circ}< {\rm RA} < 218^{\circ}$ and $-3^{\circ}< {\rm Dec} < +4^{\circ}$ with resultant RMS noise of $37~\mu$Jy/beam.
Source catalogues are publicly available at CASDA for each of the 11 tiles.
In this analysis we use three catalogues that correspond to the three tiles with the lowest RMS noise in that sky area.
We then combine these three catalogues by removing duplicate sources in the overlapping regions.
The final catalogue has a total number of 34,705 components with 3,841 complex source components.
We use source positions given in the catalogues to make cutouts from the averaged maps.

Another ASKAP survey used in the present work is the Survey With ASKAP of GAMA-09 + X-ray (SWAG-X) which as the name suggests is designed to cover the GAMA\footnote{http://www.gama-survey.org/} and eROSITA\footnote{https://www.mpe.mpg.de/eROSITA} Final Equatorial-Depth Survey \citep[eFEDS;][]{brunner21} fields.
This survey comprises 13 ASKAP tiles (publicly available at CASDA) for complete coverage of the eFEDS region, with $\sim 8$ hours integration per tile.
Similar to EMU-PS and DINGO, each tile is processed using ASKAPsoft.
The average beamwidth of the survey is $14^{\prime\prime} \times 12^{\prime\prime}$ FWHM.
The frequency band of the survey is centred at 888 MHz.
The RMS noise of these 13 tiles ranges from 49 to $64~\mu$Jy/beam.
These tiles cover the sky area with $126^{\circ}< {\rm RA} < 146^{\circ}$ and $-5^{\circ}< {\rm Dec} < +8^{\circ}$.
We generated six averaged maps by combining 2-3 tiles for each map and using weights proportional to $1/\rm RMS^2$.
The tile SBIDs used for making averaged maps include: 10132 and 20875; 10108 and 20931; 10123 and 10475; 10135 and 20132; 10126 and 21021; 10137, 10129 and 10486.
The resultant RMS noise of these six averaged maps is between 32 and $36~\mu$Jy/beam.
We use six catalogues corresponding to the tiles with the lowest RMS noise in the same sky region.
We then combine these three catalogues by removing duplicate sources in the overlapping regions.
The combined catalogue has 145,011 components with 21,324 complex source components.

As mentioned before, our analysis in this paper is limited to the complex sources from all three ASKAP surveys.
We use EMU-PS for training the ML model and the other two surveys are used to infer the trained model.
Note that the source catalogues used to get the positions of radio sources in the SWAG-X and the DINGO surveys are from the individual tiles.
However, we use the averaged maps instead of the individual tiles to make image cutouts at the positions of these radio sources.
These cutouts are then used to find the peculiar sources using the trained ML model and for the figures in the present work.
Due to lower noise in the averaged maps, it is possible that the complex radio sources detected in the individual tiles have higher signal-to-noise in the averaged maps.
Here we assume that these catalogues have all the top peculiar complex sources that are detected by our ML method.
Future work should verify this by creating source catalogues from the averaged images, which is beyond the scope of the
current work that is focused on the development of ML method from available catalogues.

\subsection{Infrared and Optical data}
\label{SEC:DES_SDSS}
We use the photometric data available for the ASKAP survey regions to identify the infrared and optical sources in the region of circular and peculiar radio objects presented here. 
The Wide-field Infrared Survey Explorer \citep[WISE;][]{wright10} is an all sky infrared survey observed in the W1, W2, W3 and W4 bands that correspond to 3.4, 4.6, 12 and 22 $\mu$m wavelength.
In this study, we use only the W1 band from AllWISE \citep[AllWISE;][]{cutri13} that has a 5$\sigma$ point source sensitivity of 28 $\mu$Jy.
The optical data were taken from the publicly available 9th data release of the Dark Energy Spectroscopic Instrument's Legacy Imaging Surveys \citep[DESI LS DR9\footnote{https://www.legacysurvey.org/dr9/};][]{schlegel21}, the Science Archive Server of Sloan Digital Sky Survey \citep[SDSS;][]{alam15} and Dark Energy Survey \citep[DES;][]{abbott18}.
Unless specified otherwise, we report photometric redshifts from the counterparts in DESI LS DR9 throughout this paper.

\section{Method}
\label{SEC:method}
The first crucial step while fitting a machine learning model is to pre-process the data and make it suitable for the machine.
In this section, we describe the pre-processing procedure as well as the machine learning technique used here.

\begin{figure*}
\centering
\includegraphics[width=18cm, scale=0.5]{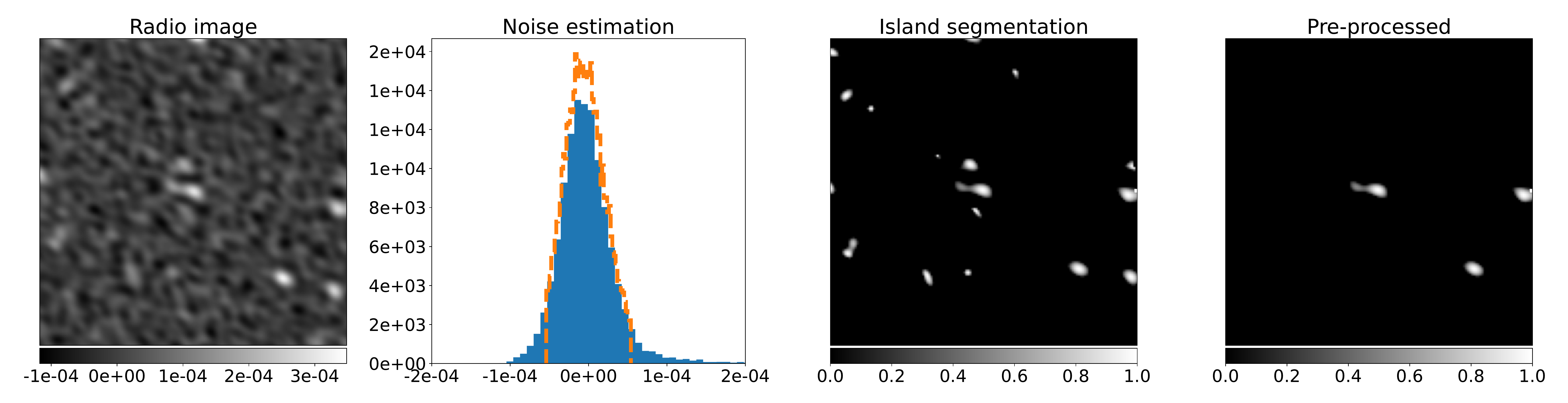}
\caption{Pre-processing procedure for radio images. From left to right, the first panel shows an ASKAP observed radio image. Second panel of Figure 1 shows the full (blue-filled histogram) and clipped (orange-dashed line) distributions of image pixels. Noise is estimated as the standard deviation ($\sigma$) of clipped distribution. Third panel shows the segmented islands at positions where pixel values are greater than 3$\sigma$. Here pixel values are converted to logarithmic scale and Min-Max normalisation is applied. Fourth panel shows the final pre-processed image where a threshold limit on number of pixels that constitute an island is imposed. This removes most of the noise fluctuations in radio maps.} 
\label{FIG:Preprocessing}
\end{figure*}

\subsection{Data Pre-processing}
\label{SEC:preprocessing}
The most important aspect of machine learning is the quality of data used to train models.
The high sensitivity of ASKAP surveys creates advanced challenges for data pre-processing due to the large source density in survey images.
We design the following pre-processing scheme to enhance useful features in the radio images:
\begin{itemize}
  \item We create cutouts from the survey images at the positions of all components of complex radio sources.
  We chose a cutout size of $5^{\prime} \times 5^{\prime}$ as only 11 sources in EMU-PS (i.e. 1 in $\sim 20,000$) have a size greater than $5^{\prime}$ \citep[][]{yew22prep}.
  This gives us a $150\times 150$ pixel image with pixel size of $2^{\prime \prime}$.
  One such cutout is shown in the left panel of Figure~\ref{FIG:Preprocessing}.
  This map has a faint double lobed radio source in the centre which has a low signal-to-noise ratio.
  \item We estimate the noise in each cutout. 
  This is done by first measuring the Median Absolute Deviation (MAD) of pixel values. 
  Two rounds of data clipping are then applied to remove outlying pixels. 
  The outlier threshold is chosen at $3\times$MAD.
  The noise is then estimated as the standard deviation of the clipped distribution.
  The second panel of Figure~\ref{FIG:Preprocessing} shows the full and clipped distributions of image pixels in blue (filled) and orange (dashed) colors.
  \item We perform an island segmentation for each cutout by generating masks of island sources with pixel values greater than 3$\sigma$.
  Here $\sigma$ is defined as the standard deviation of the clipped distribution.
  At the positions of these masks, we convert the pixel values to a logarithmic scale and perform Min-Max normalisation that enhances the signal on the scales of islands.
  The pixel values of the rest of the image are set to zero and the Min-Max normalisation of segmented regions changes the image scale in the range of 0 to 1.
  In the resultant image, shown in the third panel of Figure~\ref{FIG:Preprocessing},
  the source density is moderately high, and some of the islands may just be noise fluctuations or artefacts.
  \item To overcome this issue we impose a threshold on the number of pixels that constitute an island in the image.
  This means that we keep only those islands for which the signal is distributed over a large number of pixels.
  After some tests and visual inspections we set the minimum size for an island to 60 pixels.
  This threshold removes most of the noise fluctuations from the maps.
  Note that this limit may also remove some point sources. 
  However, that doesn't effect our analysis as the purpose of this study is to discover the most peculiar complex sources.
  The final pre-processed radio image is shown in the right panel of Figure~\ref{FIG:Preprocessing}.
\end{itemize}

\subsection{Self Organizing Map}
\label{SEC:SOM}
A self-organizing map \citep[SOM;][]{kohonen82} is a neural network that provides an efficient way to understand high-dimensional data.
The neural network constructs a representative feature map of the training dataset. 
This can be used for the tasks of dimensionality reduction and to display similarities among data sets.
SOM learns in an unsupervised manner and does not require a target vector for the dataset. 
This is important for our task as the radio sources that we aim to find are unknown objects.
An advantage of using SOM over other unsupervised architectures is topologically preserved mapping from input to output spaces.
This is important to retain the spatial information of astronomical images.

The basic unit of the SOM is a neuron $n$. 
A number of $N$ neurons are organized in an input layer and are connected to an output feature map.
These connections have associated weights $w$ that are randomly initialised.
While training, data is provided to the input layer and the extracted features are propagated to the output map.
The output map has the form of a lattice or grid where each neuron is placed at a position $p$. 
Each neuron in the lattice competes with the others to win every subject in the dataset.
For a training iteration $i$, a subject $d$ from the dataset $D$ is selected to compute a similarity measure $S(d,w_p)$ with respect to a neuron with prototype weights $w_p$.
The winning neuron for $D_j$ is its Best Matching Unit (BMU) whose position is identified as $k$.
Following this the prototype weights of BMU and neighbouring neurons are updated as
\BE
w_p^{\prime} = w_p + (\phi(d)-w_p) \times G(p,k) \times L(i),
\label{EQ:weights}
\EE
where $w_p^{\prime}$ is the updated weight. The term
$(\phi(d)-w_p)$ is required to spatially align $d$ onto $w_p$.
$G(p,k)$ is the neighbourhood function parametrised as a Gaussian that controls the propagation of weight updates to neighbouring neurons. 
In principle, the neighbouring neurons of a BMU get smaller updates and the amount depends upon the separation between $k$ and $p$ as well as the chosen width $\sigma_G$ of the Gaussian. 
$L(i)$ is the learning rate that further controls the weighting updates for each iteration.

The SOMs have been used previously in astronomy for classification of light curves and clustering of gigahertz-peaked sources \citep[e.g.][]{brett04,torniainen08}. 
More recently, SOMs are used for the estimation of photometric redshifts in large surveys \citep[e.g.][]{geach12,wright20}.
These datasets are in the form of catalogues of sources.
The application of neural networks on the astronomical image datasets requires that the method is invariant to affine transformations.
Examples of such transformations include translation, scaling, flipping and rotation of images.
This means that for sources in an image, e.g. double lobed Active Galactic Nuclei (AGN), the algorithm should not be sensitive to their orientation in the sky.
To approach a solution, \cite{ralph19} used a convolutional auto-encoder to reduce the impact of affine transformations for the classification of radio galaxies.
Similarly, \citep{segal22} is using auto-encoders to measure the complexity of radio galaxies.
However, the training of the SOM using the compressed latent vector space of auto-encoders results in the loss of topological information.
In a different approach, \cite{polsterer15} developed Parallelized rotation and flipping INvariant Kohonen maps (PINK) to incorporate the transformational invariance into the SOMs.
\cite{galvin19} showed that PINK can be an ideal solution to break the degeneracy arising due to affine transformations without losing topological information.
\cite{galvin20} further exploited PINK to classify different morphologies of radio sources using the Faint Images of the Radio-Sky at Twenty centimetres \citep[FIRST;][]{becker95}. 
Following this, we use PINK in this analysis to find the rare and unusual radio morphologies.

PINK implements a modified Euclidean distance metric for similarity measure that can be written as
\BE
S(d, w_k) = \underset{\forall \phi \in \Phi}{{\rm minimize}(\phi)} \sqrt{\sum_{c=0}^C \sum_{x=0}^X \sum_{y=0}^Y \left(w_{k(c,x,y)} - \phi(d_{c,x,y}) \right)^2},
\label{EQ:similarity}
\EE
where $\phi$ is an affine transformation drawn from a set of $\Phi$ and is optimized to align an image to features in the BMU.
This is propagated to update the neighbouring units. $C$ is the number of channels of an image.
Here we use only one channel. $X$ and $Y$ define the pixel size of the image.
This optimizes the search for transformation parameters to align $d$ to prototype weights $w_k$ of a SOM.

\begin{table}[!ht]
\centering
\begin{center}
\begin{tabular}{lccccc}
\hline
\hline
\multicolumn{1}{c}{Stage} & \multicolumn{1}{c}{Iterations} & \multicolumn{1}{c}{Rotations} & \multicolumn{1}{c}{Increments} &  \multicolumn{1}{c}{$\sigma_G$} & \multicolumn{1}{c}{$L$} \\ 
\hline
1 & 5  & 90  &$4^{\circ}$& 1.5 & 0.1  \\
2 & 5  & 180 &$2^{\circ}$& 1.0 & 0.05 \\
3 & 5  & 360 &$1^{\circ}$& 0.7 & 0.05 \\
4 & 10 & 360 &$1^{\circ}$& 0.5 & 0.005\\
\hline
\end{tabular}
\end{center}
\caption{Parameters for different stages of training. From left to right are the number of iterations, number of rotations, increment with each rotation, width of $G(p,k)$ and learning rate.}
\label{TAB:TRAINING}
\end{table}

\begin{figure}[!ht]
\centering
\includegraphics[width=8.5cm, scale=0.5]{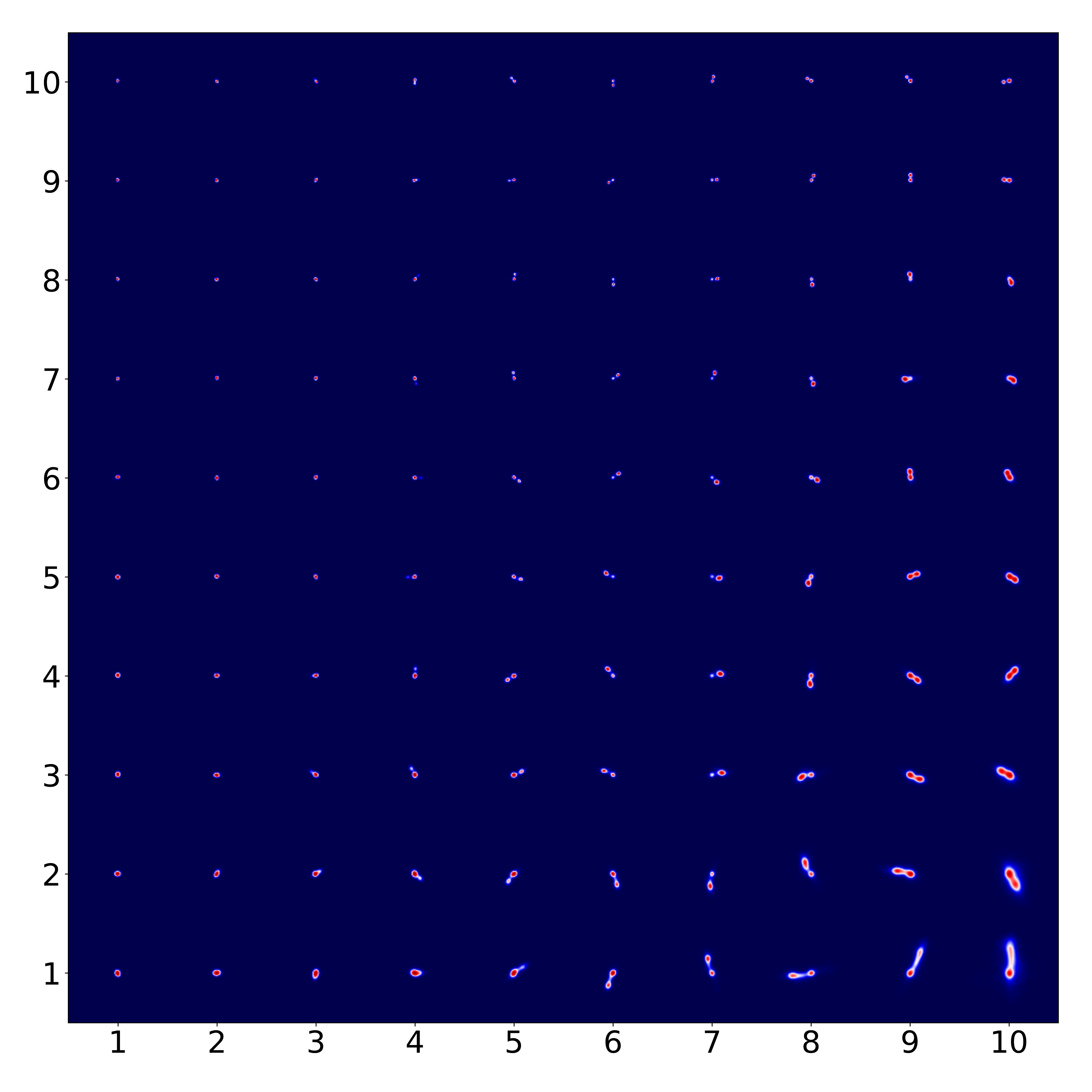}
\caption{The trained $10\times10$ SOM using the complex sources in the EMU-PS.
The X- and Y-axes show identities of neurons that are representatives of the best matched radio sources.
Across the lattice, these neurons represent resolved radio lobes, extended structures bridged by diffuse emission, and more compact sources.
This shows that after 4 stages of training, the SOM represents meaningful radio morphologies.} 
\label{FIG:SOM}
\end{figure}

\begin{figure}[!ht]
\centering
\includegraphics[width=8cm, scale=0.5]{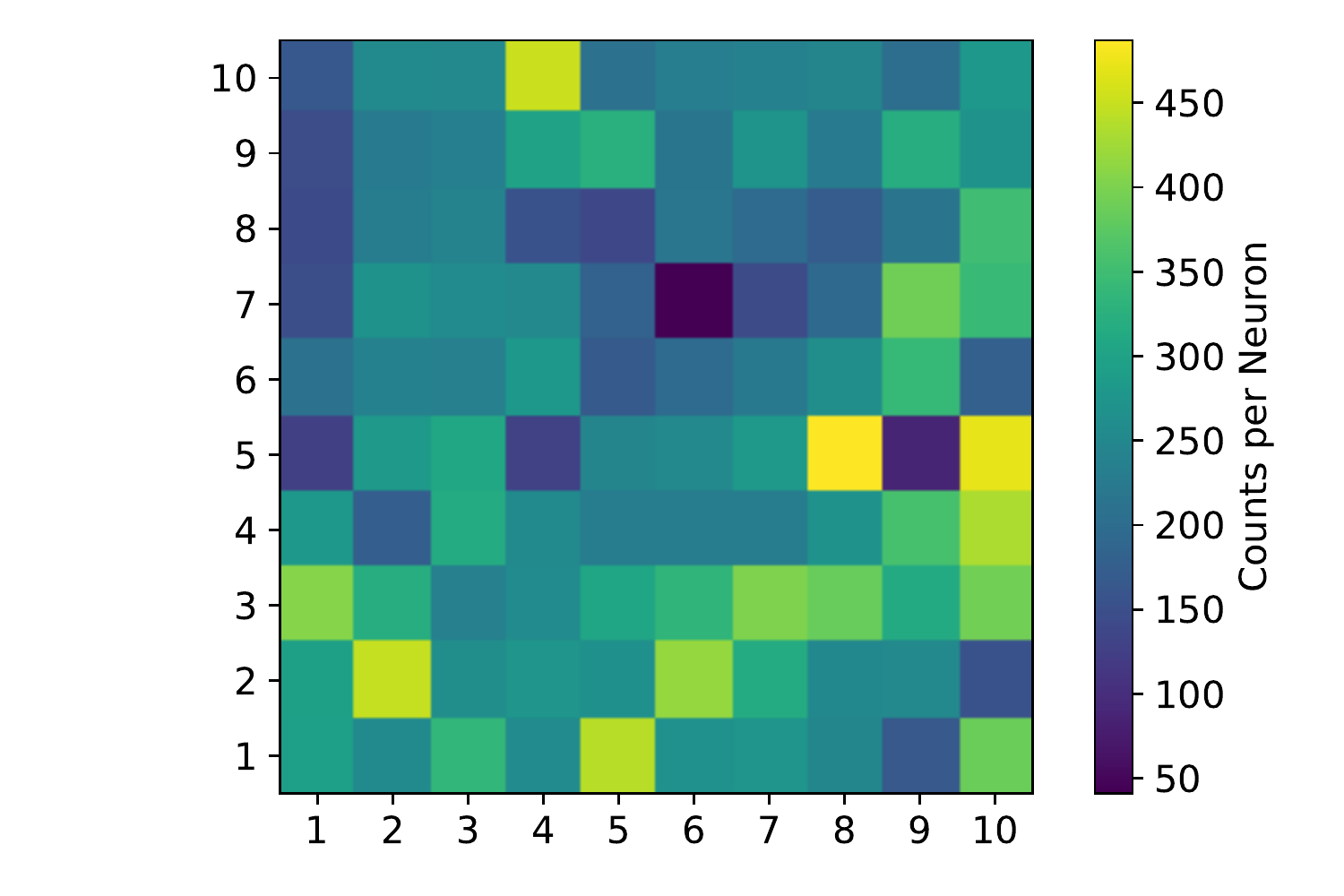}
\caption{Number counts of complex EMU-PS sources on $10\times10$ SOM lattice. 
The largest number is associated with the neuron (8,5) with resolved double lobed sources.} 
\label{FIG:BMUcounts}
\end{figure}

\subsection{Training}
\label{SEC:Training}
We construct a SOM in a Cartesian lattice space with $10\times 10$ neurons.
Each neuron has a circular shape initialised with uniform random noise between 0 and 1.
The circular shape preserves the entire region of the image against the affine transformations.
This is an improvement over the previous versions of PINK with square shaped neurons which resulted in the loss of information
in the outer regions due to image transformations \citep[e.g.][]{galvin19, galvin20}.
The SOM is trained in four stages with user-defined parameters outlined in Table~\ref{TAB:TRAINING}.

In each stage, every subject in the dataset is passed through the network to update prototype weights.
Each full passage of the dataset through the network is called an iteration.
The first three stages include five iterations each of the dataset, and the final stage has 10 iterations.
Across all stages, a normalised Gaussian neighbouring function is used to update the weights of neighbouring neurons.
The $1\sigma$ width of the Gaussian is reduced in every stage with $\sigma_G = 1.5$ and 0.5 for the first and final stages, respectively.
This helps in establishing a broad set of morphologies across the lattice in the first stage, and fine tuning of the small scale structure in later and final stages.
For the same reason, the first stage requires a minimal set of rotations.
Thus our first training stage has 90 rotations for each subject in dataset with increments of $4^{\circ}$. 
This is increased to 360 rotations in the final stage with increments of $1^{\circ}$.
The large learning rate and the size of neighbouring function in the first stage allows the modification of many prototypes with each update.
These are subsequently reduced to shrink the region of influence of each prototype weight in later stages.

Note that there are no formal convergence criteria for training a SOM as the algorithm works in an unsupervised way.
This makes the manual estimation of the training parameters an important aspect of our analysis.
With a small learning rate, the SOM will take a long computational time to train. 
On the other hand, larger values result in unstable prototype updates.
Similarly, a small neighbouring function decouples the neurons from each other, whereas its larger width results in the modification of more prototype weights.
We converge on the training parameters for the four stages by experimenting with several possibilities and qualitative examination of the meaningful morphologies across the SOM lattice.
We also train a SOM with $25\times 25$ neurons and find no difference in the detection of rare radio morphologies when compared with a SOM of $10\times 10$ neurons.

In this analysis, the SOM is trained using the 41,181 components of complex radio sources from the EMU-PS catalogue.
Each image is centred at the component position and has a cutout size of $150\times150$ pixels amounting to a $5^{\prime}\times 5^{\prime}$ field of view.
The training of the SOM is carried out on a cluster with 8 GPUs and 64 GB of memory for a total of $\sim 18$ hours.

\begin{figure}[!ht]
\centering
\includegraphics[width=8cm, scale=0.5]{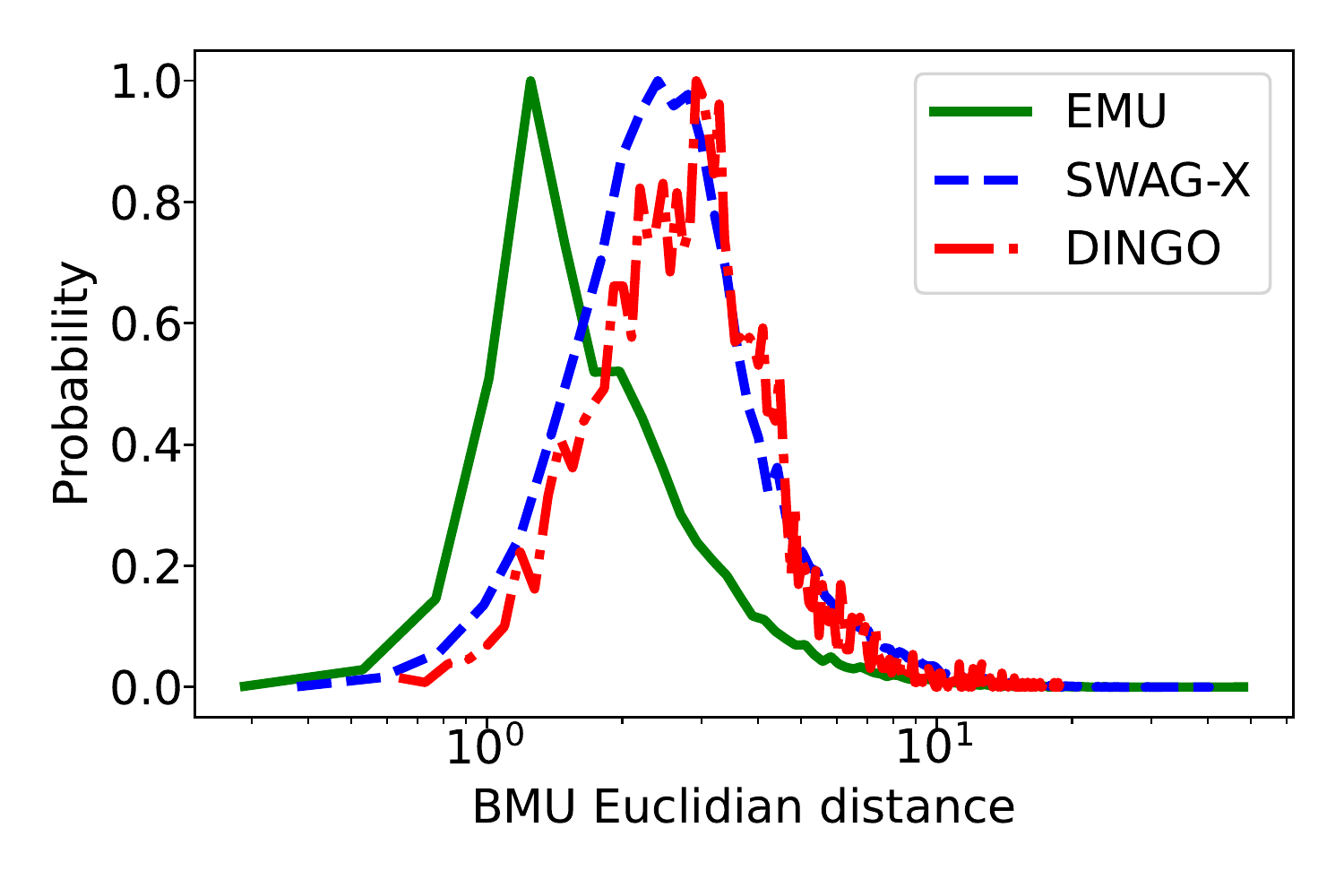}
\caption{The distributions of Euclidean distance for the EMU-PS (solid green), SWAG-X (dashed blue) and DINGO (dot-dashed red) survey datasets. 
The tails of these distributions (towards the right end) have sources among the rarest and peculiar sources (see Section~\ref{SEC:selection} for details).} 
\label{FIG:Eucl_histogram}
\end{figure}

\begin{figure*}[!ht]
\centering
\includegraphics[width=20cm, scale=0.5]{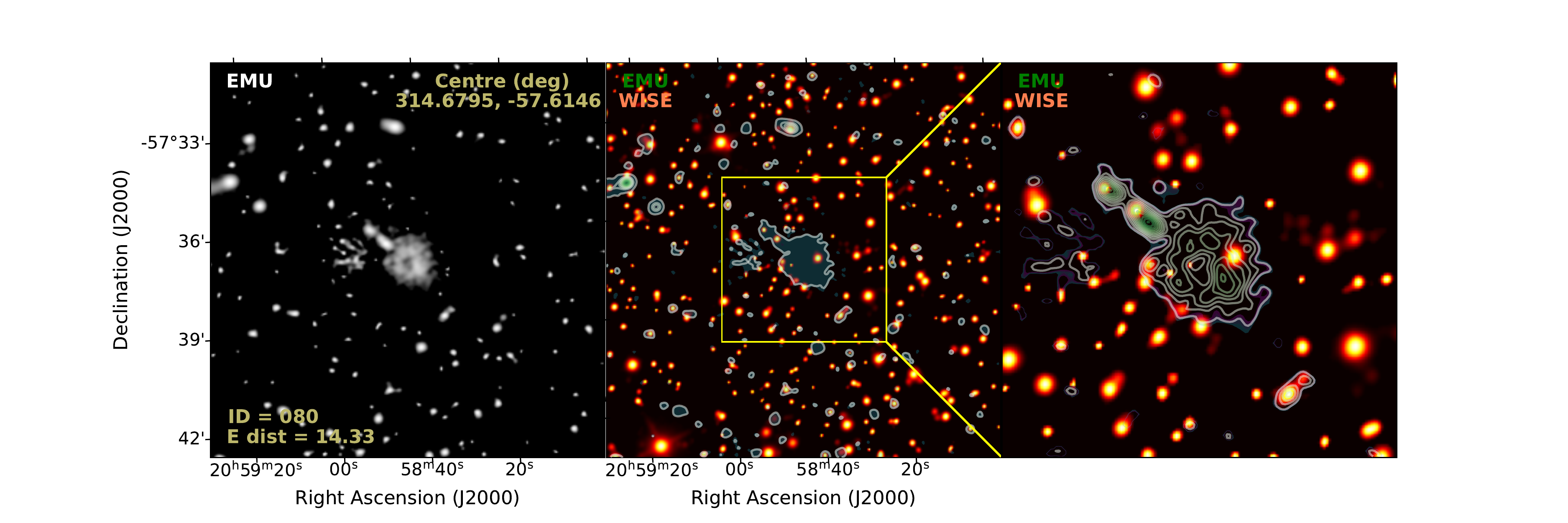}
\includegraphics[width=20cm, scale=0.5]{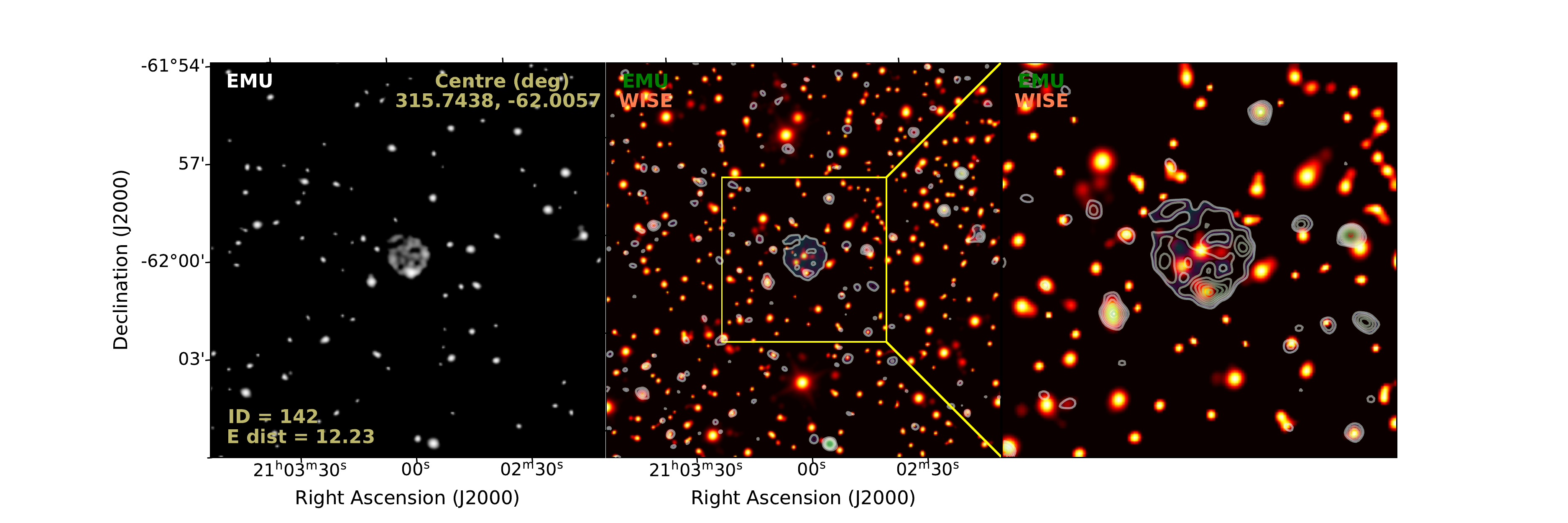}
\caption{Previously discovered ORCs located in the EMU-PS fields \citep[see][and Table~\ref{TAB:all-orcs}]{norris21b}.
The left panels show $12^{\prime} \times 12^{\prime}$ radio images from EMU-PS.
We show pre-processed radio images with no threshold on the number of pixels for an island.
The larger cutout size helps to rule out the possibility of association with other sources on large scales.
Central ORC sky positions, ID numbers for visual inspections and Euclidean distances are noted on these images.
The middle panels show radio contours on top of the WISE-W1 infrared images to visualize the nearby infrared sources.
The right panels show $5^{\prime} \times 5^{\prime}$ cutouts that is the size of the images used to train the SOM.
This shows that our method comfortably detects previously known rare morphologies among the top 0.5\% sources.
} 
\label{FIG:ORCCandidates0}
\end{figure*}

\begin{table*}[!ht]
\centering
\begin{center}
\begin{tabular}{ccccccc}
\hline
\hline
Name             &  Integrated radio    & RA (Deg).       & Dec (Deg)     & Survey       &  Reference            \\
                 &  flux density (mJy)  \\
\hline
\\
ORC J2102--6200 & 6.26                 & 315.7429       & $-62.0044$     & ASKAP        & \citet{norris21b} \\
                    &                      &                 &               & (EMU-PS)     &                       \\
ORC J2058--5736 & 6.97                 & 314.6783       & $-57.6161$     & ASKAP        & \citet{norris21b} \\
                    &                      &                 &               & (EMU-PS)     &                       \\
ORC J2058--5736 & 1.86                 & 314.7346      & $-57.6153$     & ASKAP        & \citet{norris21b} \\
                    &                      &                 &               & (EMU-PS)     &                       \\
ORC J1555+2726 & --                    & 238.8527       & $+27.4427$     & GMRT         & \citet{norris21b} \\
ORC J0102--2450 & 3.9                  & 015.6016       & $-24.8442$     & ASKAP        & \citet{koribalski21} \\
\hline
\\
J084927.5--045721 & 228.5                & 132.3645       & $-4.956 $     & ASKAP        & Present work             \\
                    &                      &                 &               & (SWAG-X)     &                       \\
J222339.5--483449 & 17.2                 & 335.9145       & $-48.5803$     & ASKAP        & Present work             \\
                    &                      &                 &               & (EMU-PS)     &                       \\
\hline
\hline
\end{tabular}
\end{center}
\caption{Previously known ORCs (top 5 rows) and ORC candidates from present work (bottom 2 rows).
From left to right we show: IDs, names using the approximated centre of diffuse emission, integrated radio flux densities, approximate geometrical centres of these systems, their parent surveys and references.}
\label{TAB:all-orcs}
\end{table*}

\subsection{Final SOM \& Selection of Rare Radio Morphologies}
\label{SEC:selection}
The final trained SOM is shown in Figure~\ref{FIG:SOM}.
After four stages of training the SOM appears to show meaningful radio morphologies.
These morphologies include resolved radio lobes, extended structures bridged by diffuse emission, and more compact sources.

The information attached to a neuron can be used to identify all subjects that share this neuron as their BMU.
A properly trained SOM contains a representative neuron for each subject in the training dataset.
Using this information, we map the image dataset on the trained SOM to evaluate the similarity statistics.
Figure~\ref{FIG:BMUcounts} shows the number counts of EMU-PS components for each of the neurons in the SOM lattice.
The lowest number of subjects in the lattice is attached to the neuron (6,7). 
The largest number is associated with the neuron (8,5) with resolved double lobed sources.
Note that SOM BMUs are representative of the majority of sources in a sample (the typical radio galaxies). Rare and unusual sources will be much more poorly characterised by the BMUs, leading to a much larger Euclidean distance than for the bulk of sources.

For an adequately trained SOM, all sources in the dataset have a BMU.
As can be noted from the prototypes in the trained SOM lattice in Figure~\ref{FIG:SOM}, all structures in the neurons can be identified as known morphologies of radio sources.
These prototypes can be used to classify these radio sources which is beyond the scope of the present work as here we are focused only on finding the rare radio morphologies.
The rare and unusual sources are not expected to be clustered in a single neuron.
Therefore, we use a similarity measure to identify the most peculiar sources in the dataset.
We use the modified Euclidean distance metric to identify these objects.
Note that the SOM is trained with EMU-PS complex sources only but we map the complex sources from all three surveys on the trained lattice.

We examine the distributions of Euclidean distances.
Figure~\ref{FIG:Eucl_histogram} shows the Euclidean distance histograms for EMU-PS (solid green), SWAG-X (dashed blue) and DINGO (dot-dashed red) complex sources.
The median (and standard deviation) of these distributions are 2.1 (2.3), 3.1 (2.4) and 3.2 (2.1) for EMU-PS, SWAG-X and DINGO, respectively.
We notice that the SWAG-X and DINGO distributions have higher median Euclidean distances as compared to the EMU-PS.
This is possibly due to the differences in observing frequencies, map resolutions and RMS sensitivities of these surveys described in Section~\ref{SEC:ASKAP} and/or a lower number of complex sources in DINGO and SWAG-X surveys.
For each of these distributions, we chose a lower limit to the Euclidean distances and visually examine the top 0.5\% of complex sources for peculiarity.
We note that this is a simplistic approach to reduce the number of visual inspections.
The choice of the top rarest 0.5\% leaves us with approximately 200, 100 and 20 sources in the EMU-PS, SWAG-X and DINGO surveys, respectively.
In the following sections, we discuss some of these rare radio source morphologies.

\floatsetup[table]{font=tiny}
\begin{table*}[!ht]
\centering
\begin{center}
\begin{tabular}{ccccccccccccc}
\hline
\hline
\\
\multicolumn{1}{c}{Name} &\multicolumn{1}{c}{RA (deg)} & \multicolumn{1}{c}{Dec (deg)} &  \multicolumn{1}{c}{Flux (mJy)} & \multicolumn{1}{c}{Counterparts} & \multicolumn{1}{c}{$g$} & \multicolumn{1}{c}{$r$} & \multicolumn{1}{c}{$i$} & \multicolumn{1}{c}{W1} & \multicolumn{1}{c}{W2} & \multicolumn{1}{c}{W1-W2} & \multicolumn{1}{c}{$z_{\rm ph}$} & \multicolumn{1}{c}{$z_{\rm spec}$} \\
\hline
\\
SWAG-X \\
J084927.5--045721\\
\\
A & 132.3638 & -4.9588  & 3  & WISEA J084927.33-045732.3  & 16.17 & 15.56 & 15.32 & 13.24 & 13.32 & -0.08 & $0.02\pm0.05$ & --\\
  &          &        &    & 2MASS J08492733-0457315    & \\
B & 132.3659 & -4.9614  & 9  & WISEA J084927.80-045741.1  & 17.78 & 16.94 & 16.39 & 12.48 & 12.45 & -0.03 & $0.08\pm0.01$ & --\\
  &          &           &    & 2MASX J08492779-0457412    & \\
C & 132.3692 & -4.9542  & 6  & WISEA J084928.60-045715.0  & 18.07 & 17.23 & 16.81 & 12.54 & 12.49 & -0.05 & $0.08\pm0.02$ & 0.07697\\
  &       &            &    & 2MASX J08492860-0457152    & \\
D & 132.3684 & -4.9505  & 18 & WISEA J084928.42-045702.1  & 18.36 & 17.48 & 17.01 & 12.69 & 12.69 & 0.00  & $0.08\pm0.01$ & --\\
  &       &            &    & 2MASS J08492840-0457017    & \\
E & 132.3607 & -4.9544  & 2  & WISEA J084926.56-045715.9  & 18.85 & 18.1  & 17.68 & 14.34 & 14.31 & 0.03  & $0.09\pm0.01$ & --\\

\\
\hline
\\
EMU-PS \\
J222339.5--483449\\
\\
A & 335.9158 & -48.5827 &0.06& WISEA J222339.73-483457.9  & 20.78 & 19.35 & 18.87 & 15.71 & 15.45 & 0.26 & $0.34\pm0.04$ & --\\
B & 335.9148 & -48.5903 &0.10& WISEA J222339.53-483524.8  & 18.76 & 17.61 & 17.19 & 15.05 & 14.77 & 0.28 & $0.22\pm0.02$ & --\\
  &            &        &    & 2MASS J22233951-4835247    & \\
C & 335.9145 & -48.5803 &0.06& WISEA J222343.07-483440.6  & 19.51 & 18.32 & 17.93 & 14.52 & 14.17 & 0.35 & $0.23\pm0.01$ & --\\
  &            &        &    & 2MASS J22234313-4834406    & \\
D & 335.9075 & -48.5785 &0.07& WISEA J222337.80-483442.4  & 21.27 & 19.94 & 19.41 & 15.78 & 15.52 & 0.26 & $0.33\pm0.04$ & --\\
\\
\hline
\hline
\end{tabular}
\end{center}
\caption{Properties of optical and infrared sources near the two new ORC candidates presented in the present work.
From left to right, we show ORC names and prominent optical sources.
Right Ascension (RA) and Declination (Dec) of these sources.
Integrated radio flux density estimated at their positions using ASKAP images.
The optical ($gri$) and infrared (W1, W2) photometry for each of the nearby sources.
Photometric redshifts from DESI LS DR9 and spectroscopic redshifts where available.
The $gri$ information for SWAG-X J084927.5--045721 is taken from Pan-STARRS \citep{flewelling20} and for EMU-PS J2223-4834 from DES surveys.
W1, W2 band information is from the WISE survey.
Photometric redshifts are taken from DESI LS DR9.
}
\label{TAB:ORC-counterparts}
\end{table*}

\begin{figure*}[!ht]
\centering
\includegraphics[width=20cm, scale=0.5]{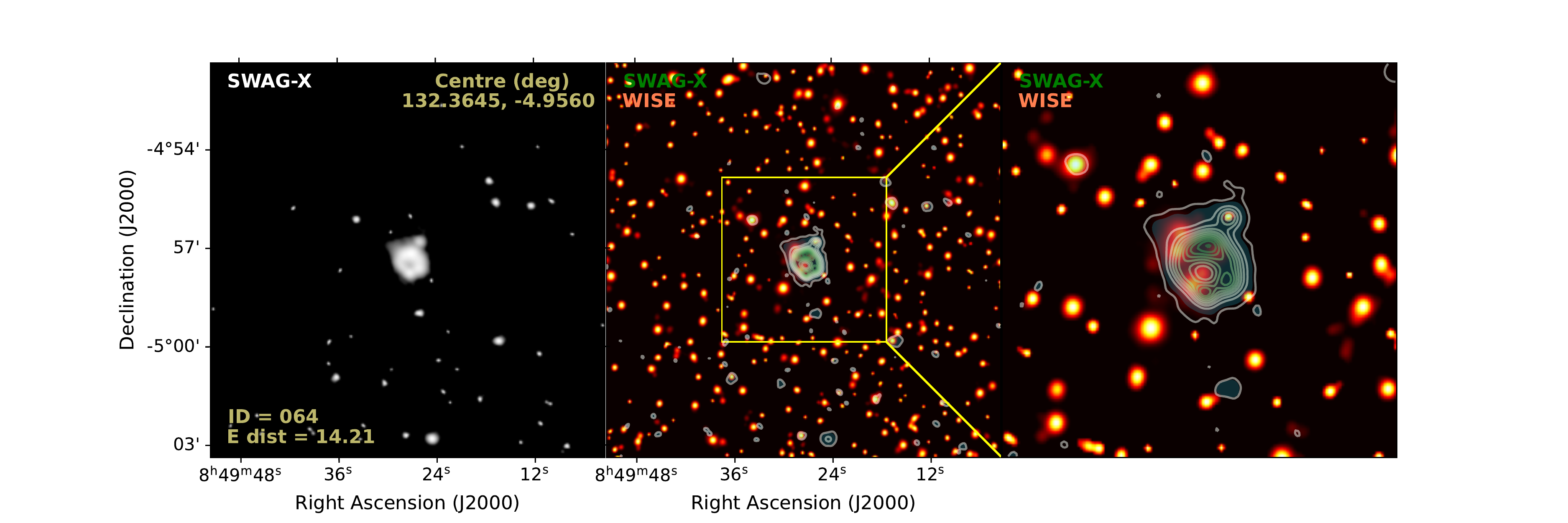}
\includegraphics[width=20cm, scale=0.5]{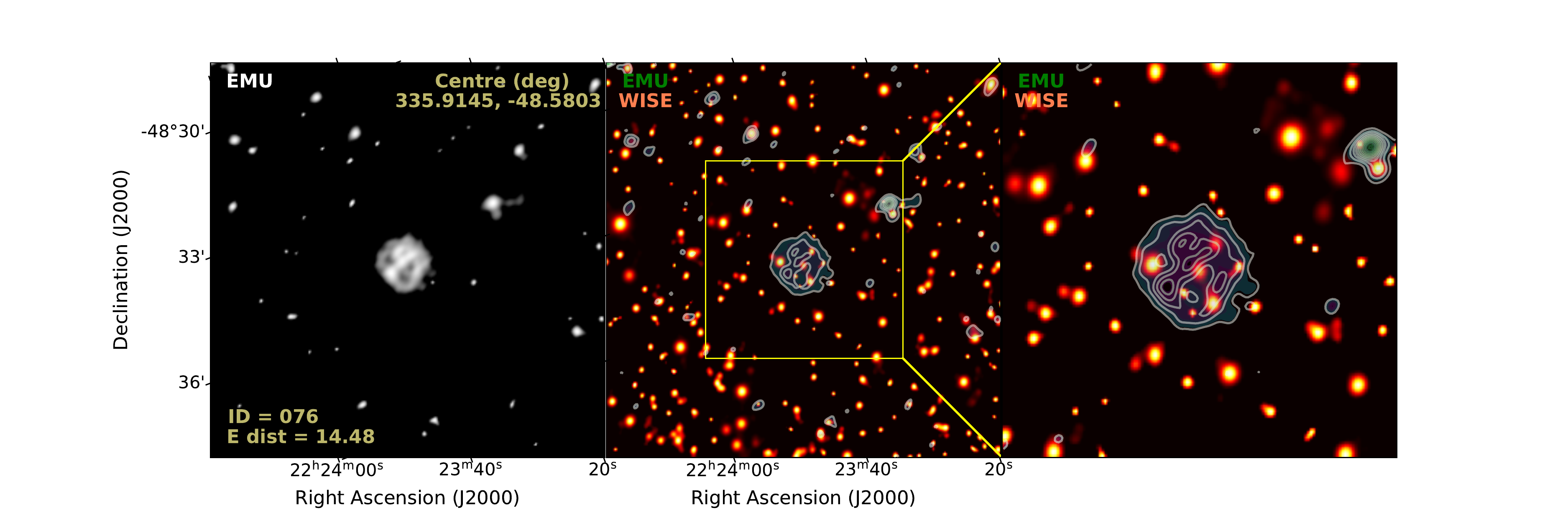}
\caption{ORC candidates from present work: SWAG-X J084927.5--045721 (top panels) and EMU-PS J222339.5--483449 (bottom panels).
Radio continuum images (left panels), radio contours overlaid on WISE-W1 infrared images (middle panels), and smaller cutouts (right panels).
Left and middle panels have a size of $12^{\prime} \times 12^{\prime}$ and right panels show $5^{\prime} \times 5^{\prime}$ cutouts that is the same size used to train the SOM.
Left panels show central sky positions, ID numbers for visual inspections and Euclidean distances noted on the images.} 
\label{FIG:ORCCandidates1}
\end{figure*}

\begin{figure*}
\centering
\includegraphics[width=8.5cm, scale=0.5, trim = 0cm 5cm 0cm 5.8cm]{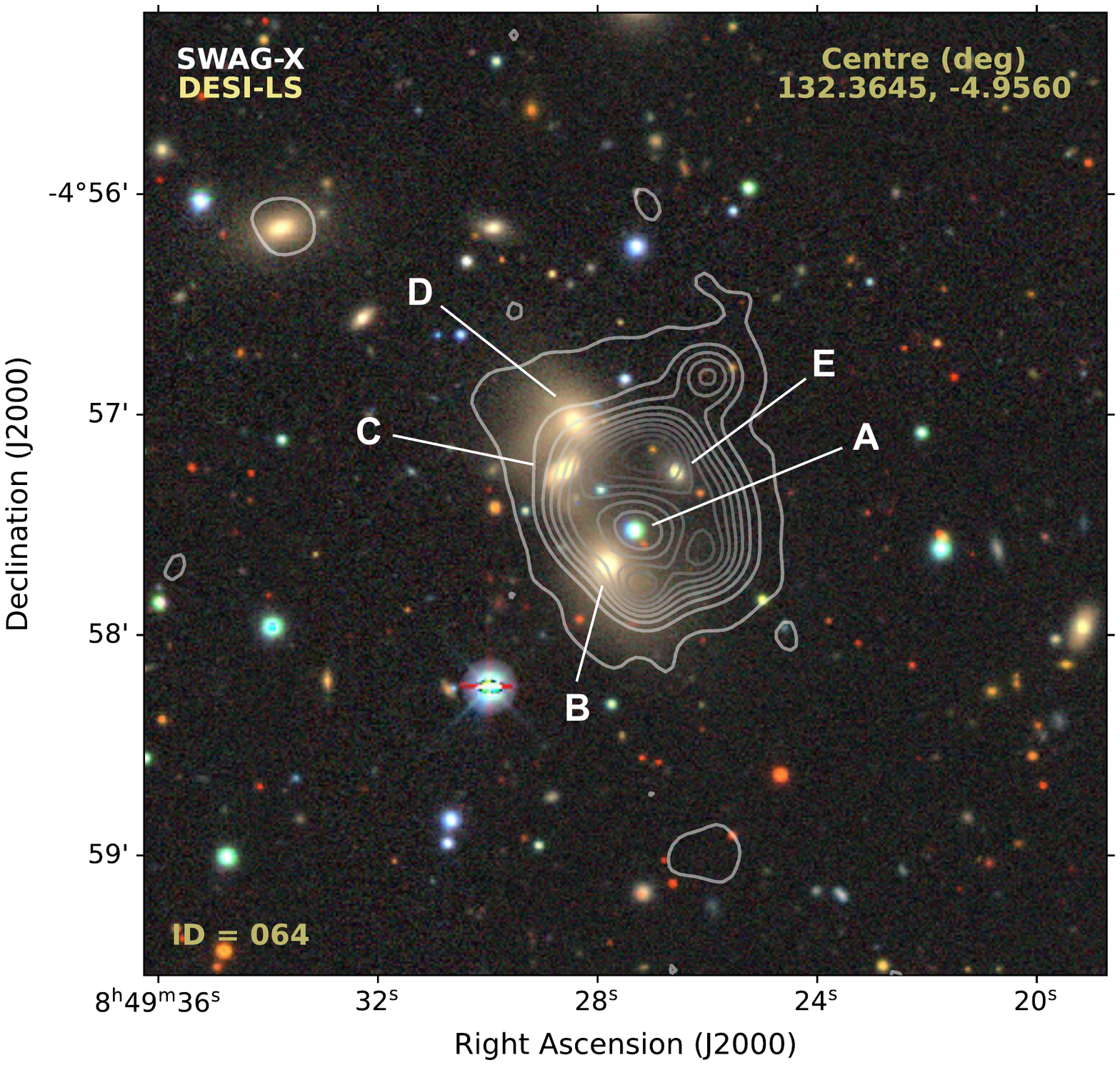}
\includegraphics[width=8.5cm, scale=0.5, trim = 0cm 5cm 0cm 5.8cm]{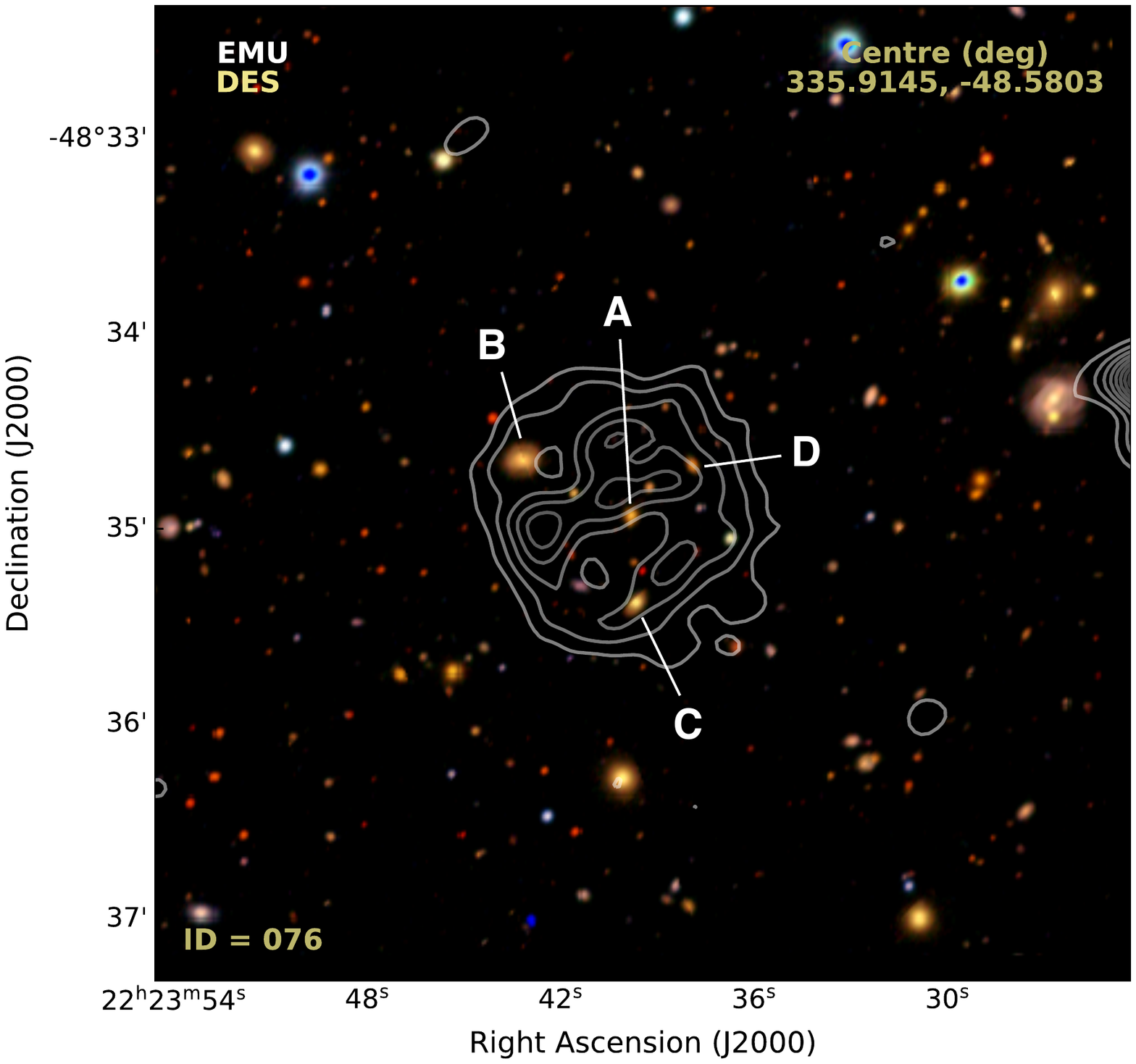}
\caption{Radio continuum contours overlaid on optical 3-color composite image ($5^{\prime} \times 5^{\prime}$ cutouts).
Optical image from DESI LS DR9 is used for SWAG-X J084927.5--045721 (left panel) and DES image for EMU-PS J222339.5--483449 (right panel).
Several optical/infrared sources are identified near each ORC candidate with counterparts in WISE and 2MASS surveys and are labelled in alphabetical order (see Table~\ref{TAB:ORC-counterparts} also).
}
\label{FIG:ORCs2}
\end{figure*}

\section{Results}
\label{SEC:results}
In this section, we present peculiar radio source morphologies among the top 0.5\% complex sources along with their observations in optical and infrared bands.
The peculiar radio sources have unconventional shapes with no corresponding diffuse emissions at optical wavelenghts.
Note that the purpose of this study is to streamline the detection of rare radio morphologies using machine learning.
Future work should should study each of these in more detail to uncover the mechanisms of their formation.
In addition to peculiar sources, we discuss examples of other conventional radio morphologies among the top 0.5\% complex sources.

\subsection{Peculiar Radio Morphologies}
\label{SEC:ORCs}
Among the peculiar radio morphologies, we find sources with nearly circular diffuse radio emission.
Such circular shapes are well known in radio images, and they either arise due to imaging artefacts or are real physical structures.
Among the known circular structures are the supernova remnants, planetary nebulae, circumstellar shells, face-on spiral galaxies or protoplanetary discs.
In a recent study, \cite{norris21b} reports the discovery of a new class of circular features in radio images and named them as Odd Radio Circles (ORCs).
They report the discovery of three ORCs in EMU-PS and one in archival data from the Giant Metrewave Radio Telescope \citep[GMRT;][]{ananthakrishnan01}.
Another ORC was discovered by \cite{koribalski21} using a different ASKAP survey.
All of these are identified serendipitously by visual inspection of the radio images (see Table~\ref{TAB:all-orcs} for the complete list).
Three out of these five  previously discovered ORCs have a central galaxy.

Figure~\ref{FIG:ORCCandidates0} shows two of the previously discovered ORCs in EMU-PS \citep[ORC J2102–6200 and ORC J2058-5736;][]{norris21b}, 
and our method places them among the top 0.5\% complex sources.
Each row in the figure has three panels.
The left panels show radio images of $12^{\prime} \times 12^{\prime}$ size.
Throughout the present work, we show pre-processed radio images with no threshold on the number of pixels for an island.
Central pixel sky positions, ID numbers for visual inspections and Euclidean distances are noted on these images.
The value of ID increases with decreasing Euclidean distance and describes the chronology for visual inspections in order of decreasing complexity. 
For example, a source with ID $=0$ is termed most peculiar with highest Euclidean distance and has highest priority for visual inspection.
The maximum value for the ID is equivalent to the number of top 0.5\% complex sources.
The middle panels show same sized infrared images from WISE W1 bands on top of the radio contours.
The larger images show that there are no prominent structures near the ORCs to which these objects may have possible associations (see Section~\ref{SEC:familiar} for other examples).
The right panels show smaller cutouts of $5^{\prime} \times 5^{\prime}$, the size used to train the SOM, with radio contours overlaid on the infrared image.

In this paper, we present two more ORC candidates that are also among the top 0.5\% sources and are similar to other previously known ORCs.
Table~\ref{TAB:all-orcs} presents positions and integrated flux densities of previously known ORCs and two ORC candidates from this analysis. 
These positions correspond to their approximate geometrical center.
Table~\ref{TAB:ORC-counterparts} shows the properties of infrared and optical sources within the extent of the continuum emission of these ORC candidates.
We present positions, ASKAP fluxes, counterparts in different surveys, redshifts, and types of morphology from literature.
The $gri$ colors and WISE (W1, W2) photometry are also shown.

\begin{figure*}
\centering
\includegraphics[width=20cm, scale=0.5]{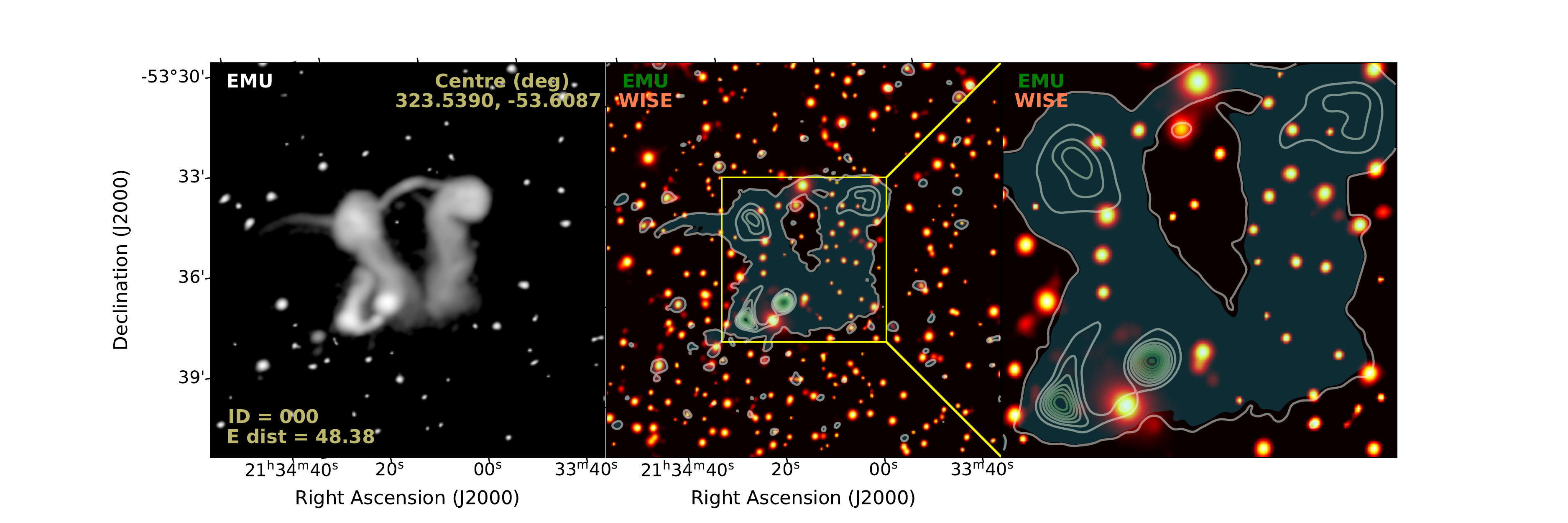}
\includegraphics[width=20cm, scale=0.5]{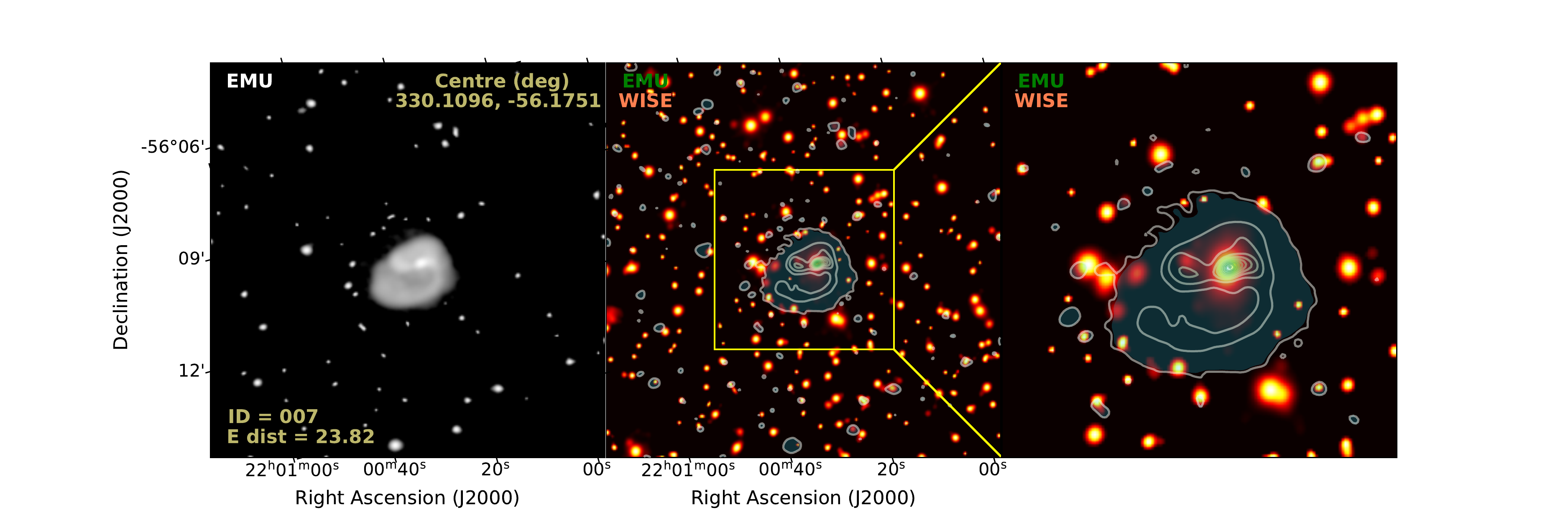}
\includegraphics[width=20cm, scale=0.5]{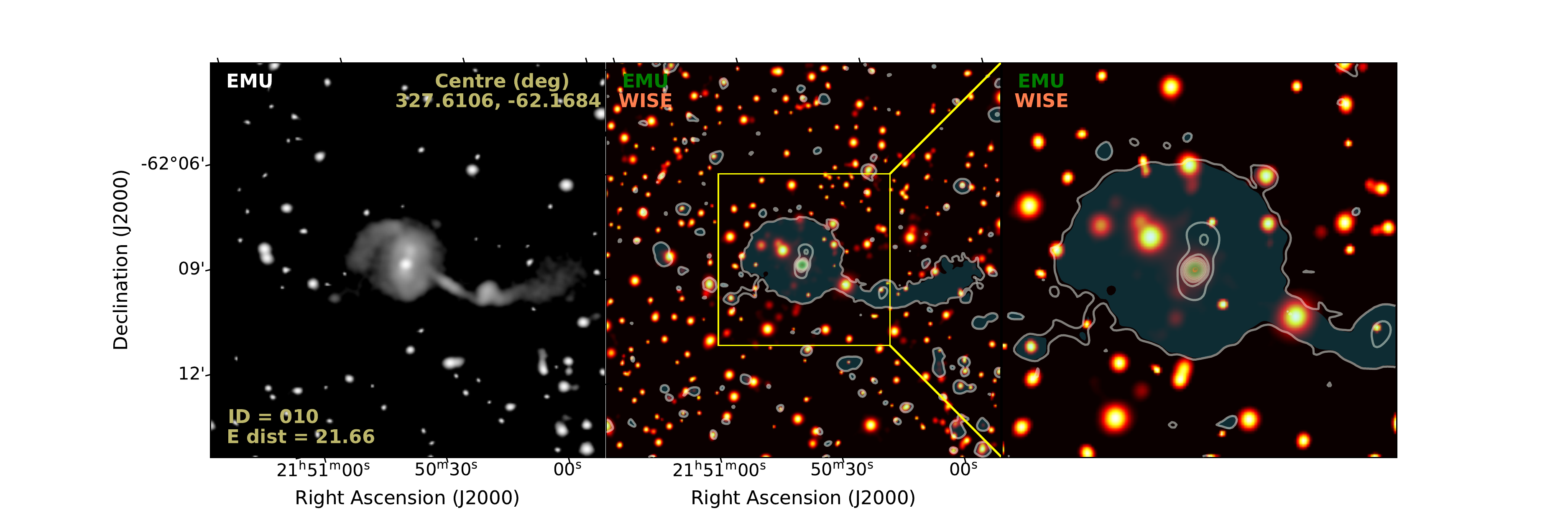}
\caption{Peculiar radio morphologies in EMU-PS: Radio morphologies other than the ORCs and among the top rarest 0.5\% of sources selected for visual inspections.
From top to bottom we show three radio sources namely EMU-PS J213409.5--533631, EMU-PS J220026.3--561030 and EMU-PS J215026.5--621006.
The description of the panels is same as Figure~\ref{FIG:ORCCandidates0}.
Both left and middle panels are $12^{\prime} \times 12^{\prime}$ large and right panels are of the same size that is used to train the SOM ($5^{\prime} \times 5^{\prime}$). 
} 
\label{FIG:unusual_radio_shapes-1}
\end{figure*}

\begin{figure*}[!ht]
\centering
\includegraphics[width=20cm, scale=0.5]{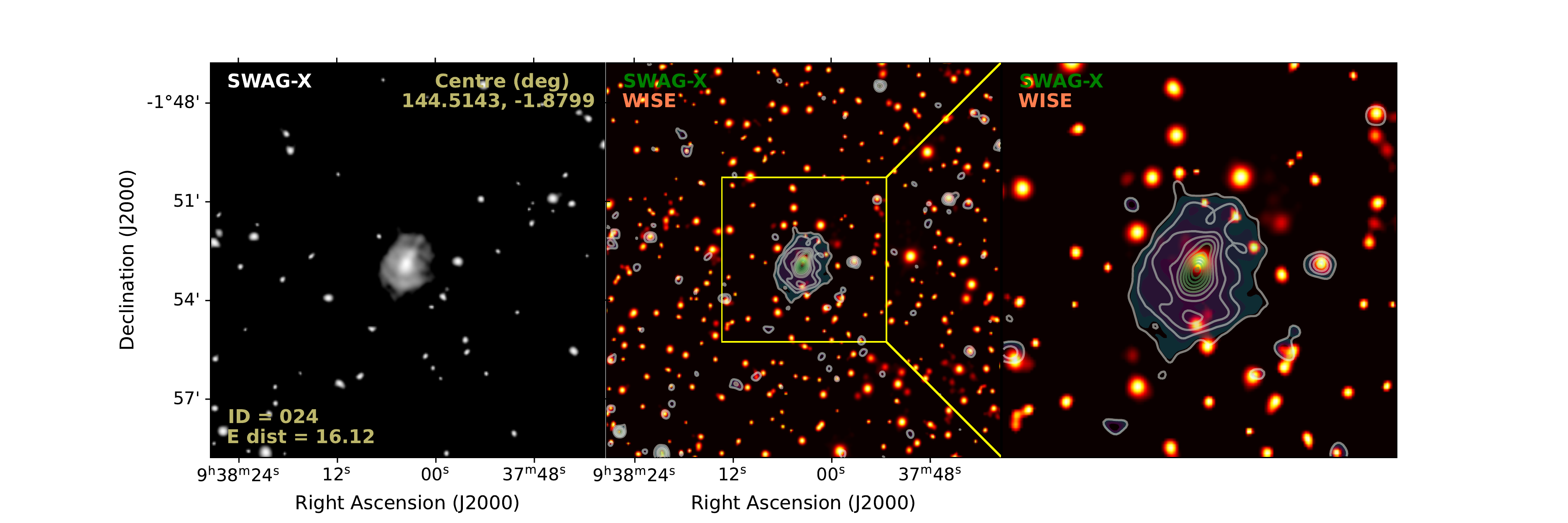}
\includegraphics[width=20cm, scale=0.5]{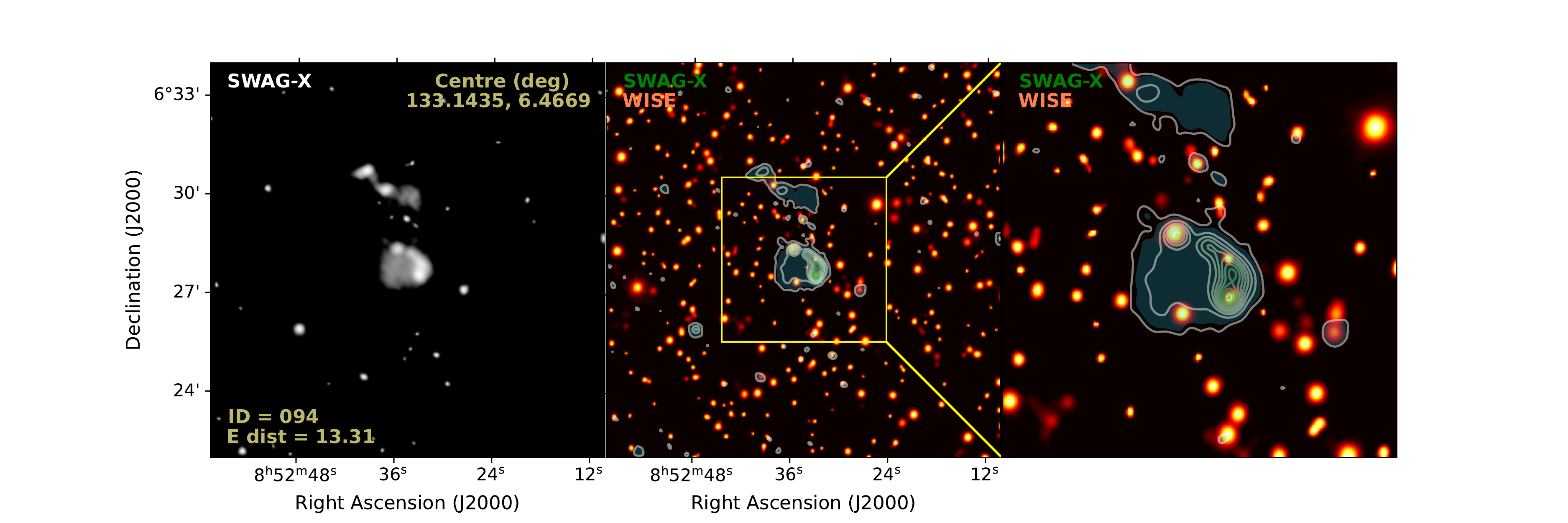}
\caption{Peculiar radio morphologies in SWAG-X: Radio morphologies other than the ORCs and among the 0.5\% sources selected for visual inspections.
From top to bottom we show two radio sources namely SWAG-X J093803.4--015247 and SWAG-X J085234.4+062801.
The description of the panels is same as Figure~\ref{FIG:ORCCandidates0}.
Both left and middle panels are $12^{\prime} \times 12^{\prime}$ large and right panels are of the same size that is used to train the SOM ($5^{\prime} \times 5^{\prime}$).See Figure~\ref{FIG:unusual_radio_shapes-1-3big} and Table~\ref{TAB:PEC-counterparts} as well.} 
\label{FIG:unusual_radio_shapes-2}
\end{figure*}

\begin{figure*}
\includegraphics[width=8.cm, scale=0.5, trim = 0cm 4.5cm 0cm 7.cm]{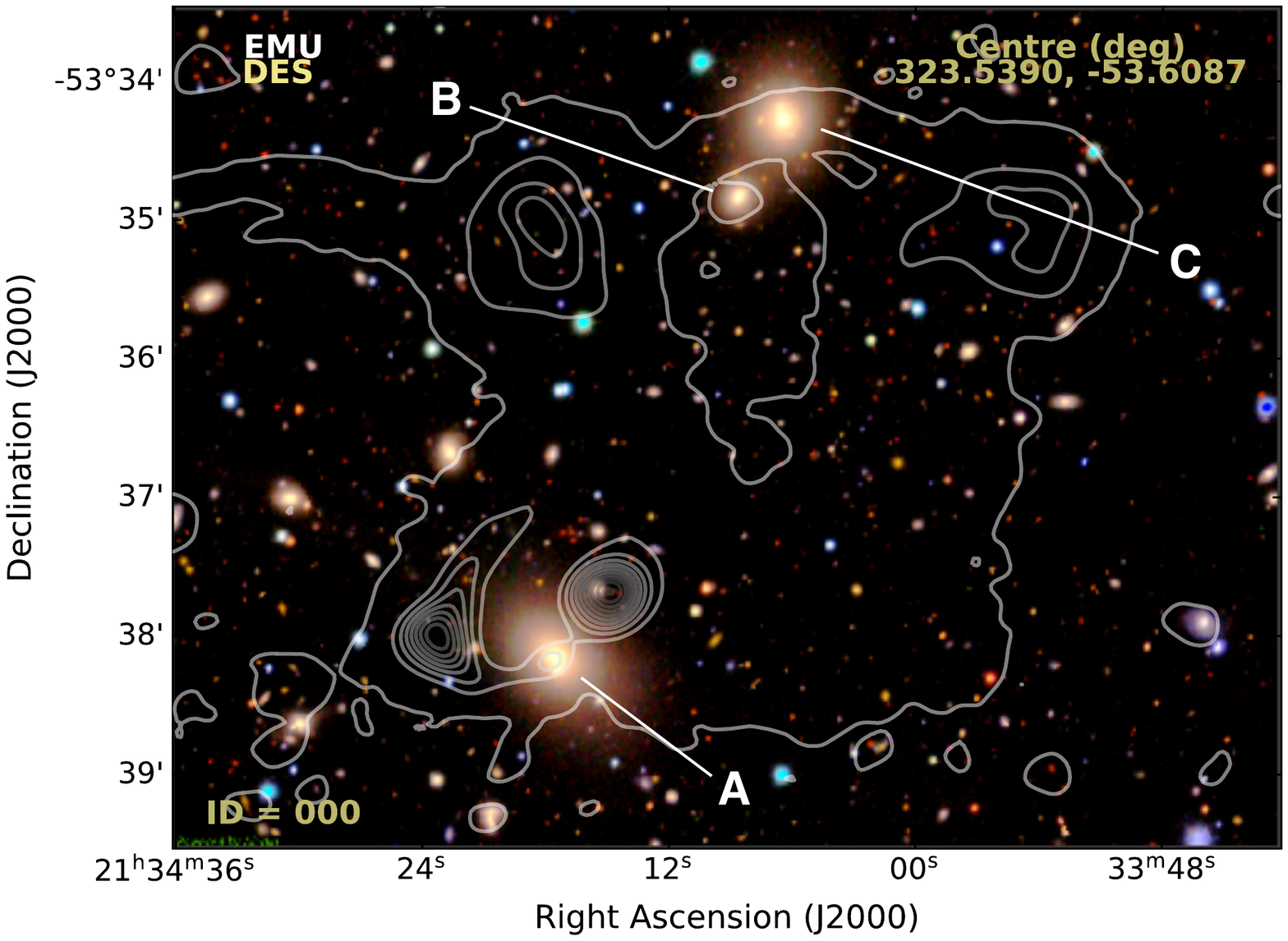}
\includegraphics[width=8.cm, scale=0.5, trim = 0cm 5cm 0cm 9.8cm]{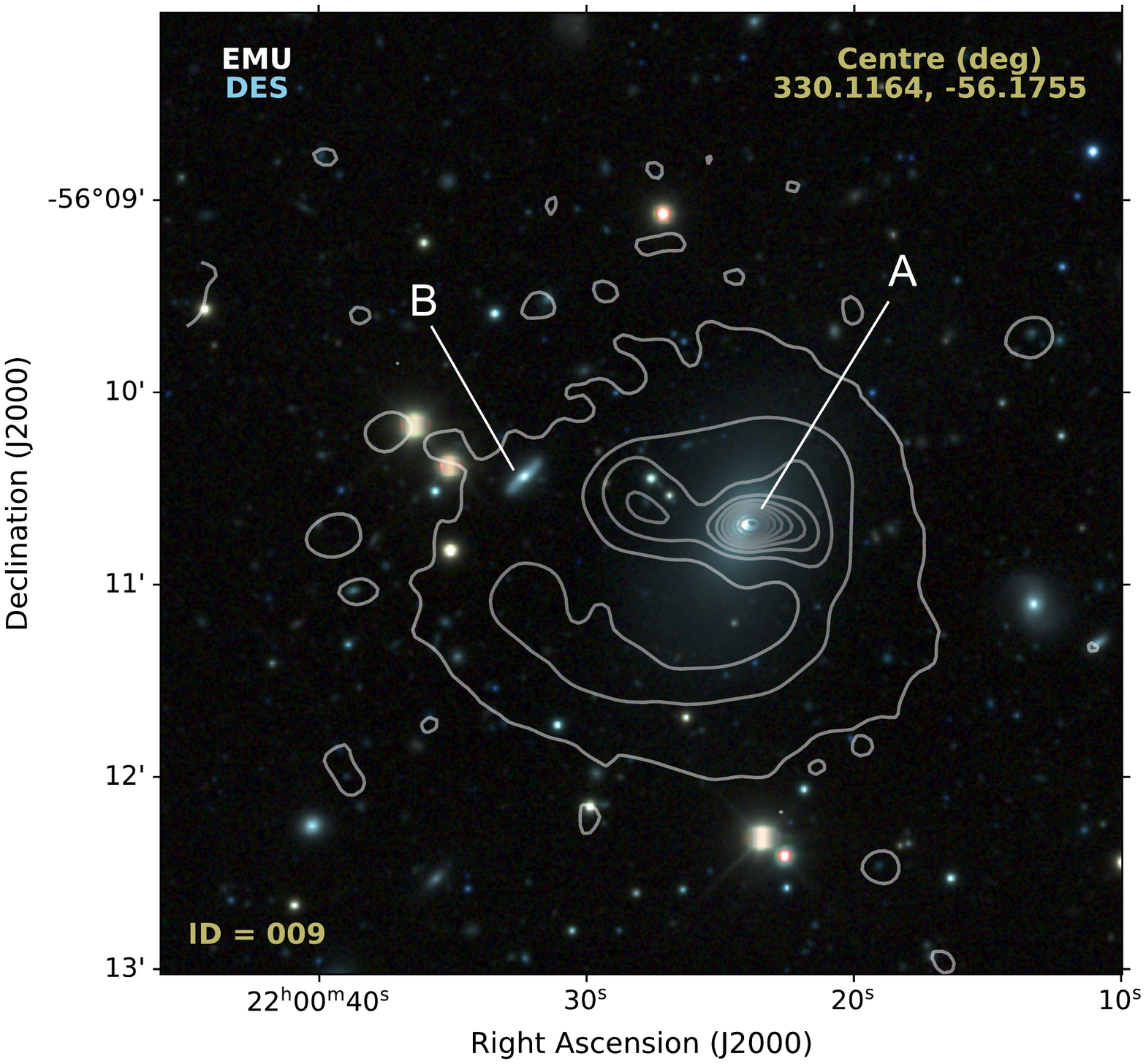}
\includegraphics[width=8.cm, scale=0.5, trim = 0cm 5cm 0cm 5.8cm]{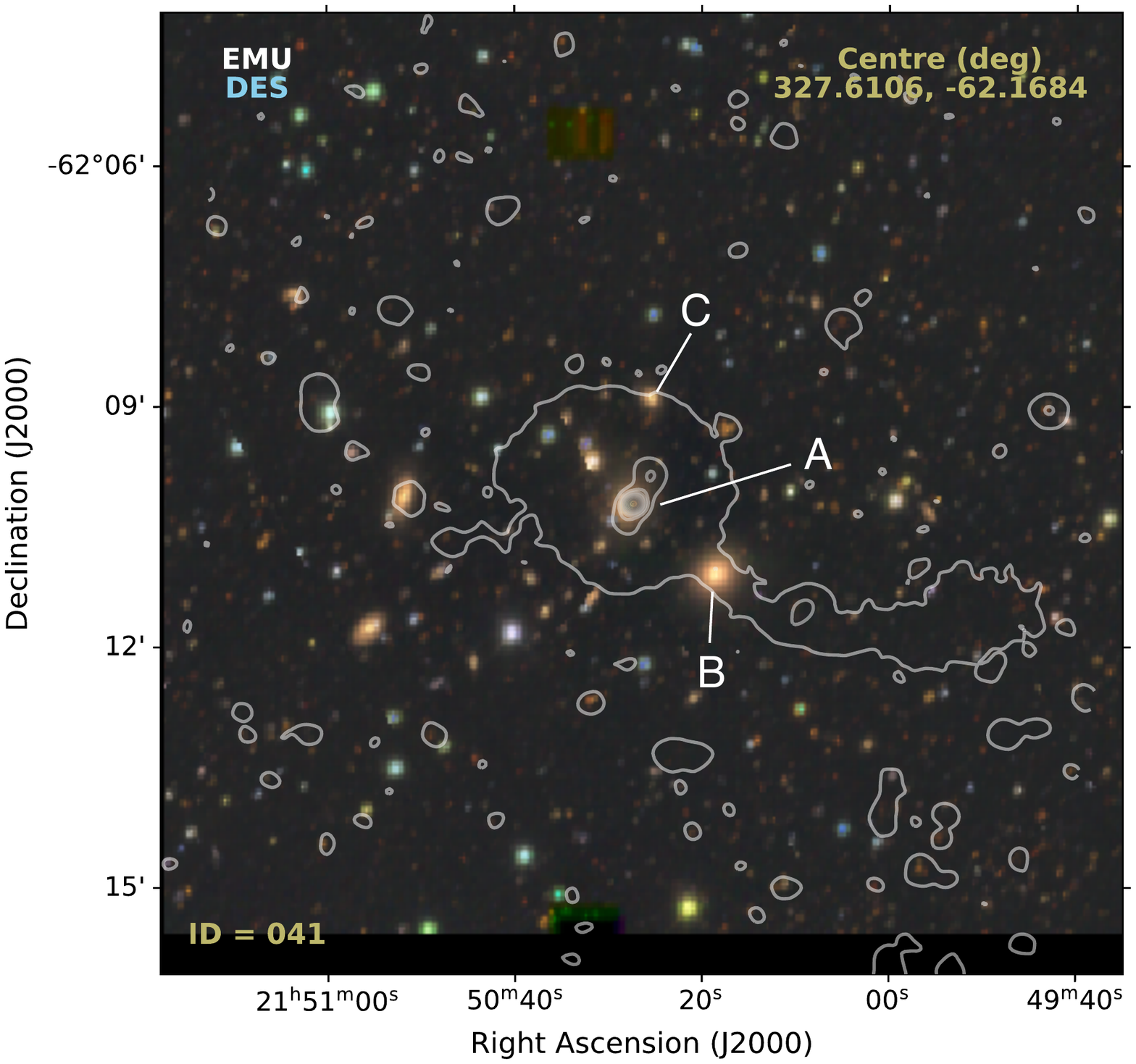}
\includegraphics[width=8.cm, scale=0.5, trim = 0cm 5cm 0cm 5.8cm]{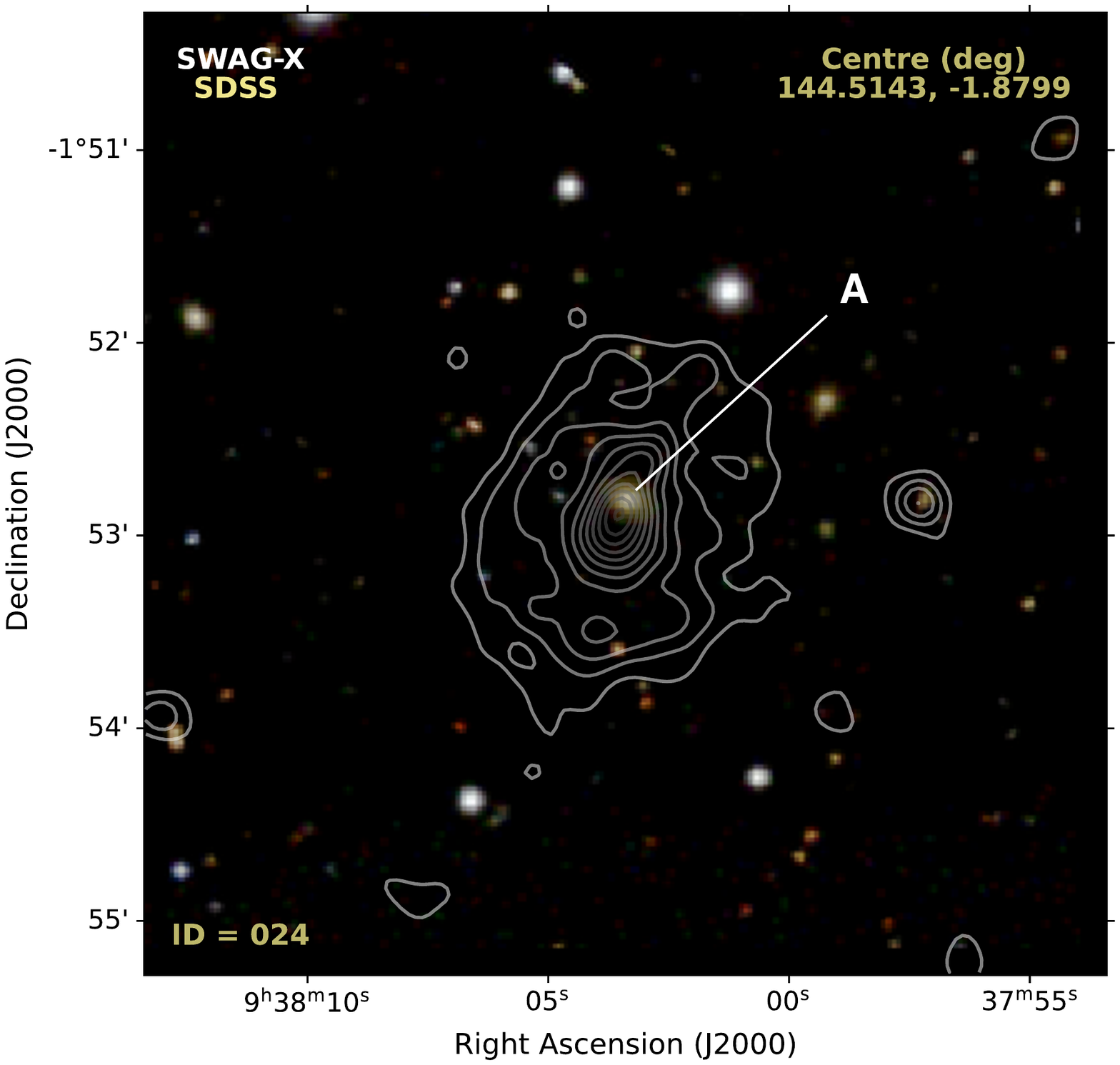}
\includegraphics[width=8.cm, scale=0.5, trim = 0cm 5cm 0cm 5.8cm]{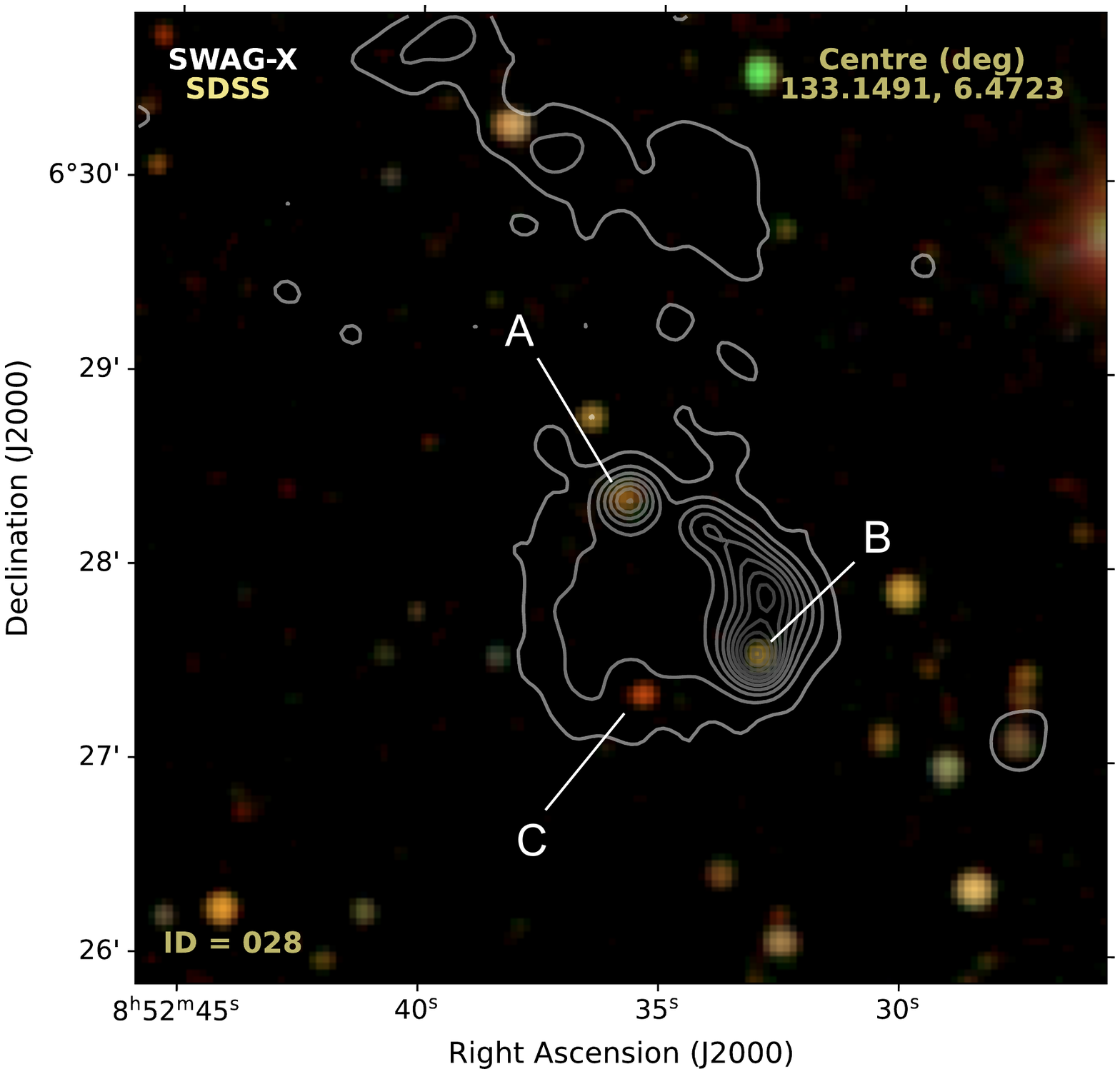}
\caption{Peculiar radio morphologies in EMU-PS and SWAG-X: Panels show radio continuum contours overlaid on DES and SDSS 3-color ($gri$) composite images.
The peculiar sources are EMU-PS J213409.5--533631 (top left), EMU-PS J220026.3--561030 (top right), EMU-PS J215026.5--621006 (middle left), SWAG-X J093803.4--015247 (middle right) and SWAG-X J085234.4+062801 (bottom).
We identify optical/infrared sources near the radio emission for each source labelled with capital letters (see details in Table~\ref{TAB:PEC-counterparts}).} 
\label{FIG:unusual_radio_shapes-1-3big}
\end{figure*}

\begin{table*}[!ht]
\centering
\begin{center}
\begin{tabular}{ccccccccccccc}
\hline
\hline
\\
\multicolumn{1}{c}{Name} &\multicolumn{1}{c}{RA (deg)} & \multicolumn{1}{c}{Dec (deg)} &  \multicolumn{1}{c}{Flux (mJy)} & \multicolumn{1}{c}{Counterparts} & \multicolumn{1}{c}{$g$} & \multicolumn{1}{c}{$r$} & \multicolumn{1}{c}{$i$} & \multicolumn{1}{c}{W1} & \multicolumn{1}{c}{W2} & \multicolumn{1}{c}{W1-W2} & \multicolumn{1}{c}{$z_{\rm ph}$} & \multicolumn{1}{c}{$z_{\rm spec}$} \\ 
\hline
\\
EMU-PS \\
J213409.5--533631 \\
\\
A     & 323.5738 & -53.6363   &18  & WISEA J213417.69-533811.1  & 15.24 & 14.29 & 13.90 & 11.49 & 11.48 & 0.01  & $0.07\pm0.03$ & 0.0763\\
      &            &              &    & 2MASX J21341775-5338101    & \\
B     & 323.5367 & -53.5811    &5.4 & WISEA J213408.81-533451.8  & 16.39 & 15.44 & 15.07 & 12.75 & 12.73 & 0.02  & $0.11\pm0.06$ & --\\
      &            &              &    &2MASX J21340880-5334516     & \\
C     & 323.5278 & -53.5719    &0.4 & WISEA J213406.70-533418.7  & 15.29 & 14.35 & 13.97 & 11.74 & 11.71 & 0.03  & $0.08\pm0.01$ & 0.07836\\
      &            &              &    & 2MASX J21340666-5334186    & \\
\\
\hline
\\
EMU-PS \\
J220026.3--561030 \\
\\
A      & 330.1004 & -56.1782  &110 & WISEA J220024.11-561041.7 & 14.93 & 13.99 & 13.59 & 11.71 & 11.75 & -0.04 & $0.05\pm0.01$ & 0.0757\\
       &            &             &    & 2MASX J22002408-5610413   & \\
B      & 330.1346 & -56.1742  &1   & WISEA J220032.19-561026.0 & 17.20 & 16.25 & 15.86 & 13.59 & 13.58 & 0.01  & $0.08\pm0.01$ & --\\
       &            &             &    & 2MASX J22003234-5610273   & \\
\\
\hline
\\
EMU-PS \\
J215026.5--621006 \\
\\
A      & 327.6138 & -62.1703  & 36 & WISEA J215027.29-621013.3 & 15.79 & 14.85 & 14.47 & 12.10 & 12.08 & 0.02  & $0.07\pm0.01$ & -- \\
       &            &             &    & 2MASX J21502732-6210129   & \\
B      & 327.5745 & -62.1852  & 4  & WISEA J215017.86-621106.4 & 15.66 & 14.77 & 14.39 & 12.28 & 12.27 & 0.01  & $0.06\pm0.01$ & --\\
       &            &             &    & 2MASX J21501790-6211070   & \\
C      & 327.6038 & -62.1485  & 3  & WISEA J215024.94-620854.5 & 17.22 & 16.28 & 15.90 & 13.37 & 13.33 & 0.04  & $0.08\pm0.01$ & --\\
       &            &             &    & 2MASX J21502489-6208550   & \\
\\
\hline
\\
SWAG-X\\
J093803.4--015247\\
\\
A & 144.5139 & -1.88          & 2  & WISEA J093803.35-015247.9 & 18.44 & 17.06 & 16.56 & 13.94 & 13.65 & 0.29  & $0.22\pm0.01$ & --\\
  &            &                  &    & 2MASS J09380334-0152480   & \\
\\
\hline
\\
SWAG-X \\
J085234.4+062801 \\
\\
A      & 133.149  & 6.4725    & 2.1& WISEA J085235.74+062821.1 & 18.54 & 17.35 & 13.59 & 13.99 & 13.51 & 0.48  & $0.19\pm0.02$ & 0.15958\\
       &            &             &    & 2MASX J08523573+0628209   & \\
B      & 133.1357 & 6.4605    & 1.2& WISEA J085232.90+062731.9 & 21.84 & 20.8  & 20.49 & 15.55 & 15.8  & -0.25 & $0.18\pm0.06$ & --\\
       &            &             &    & SDSS J085232.91+062731.7 & \\
C      & 133.1442 & 6.455     & 0.3& WISEA J085235.31+062720.2 & 22.1  & 20.93 & 20.32 & 13.92 & 13.78 & 0.14  & $0.26\pm0.05$ & --\\
       &            &             &    & SDSS J085235.32+062720.5 & \\
\\
\hline
\hline
\end{tabular}
\end{center}
\caption{Properties of optical and infrared sources near the peculiar radio sources other than the ORC candidates.
The columns are the same as described in Table~\ref{TAB:ORC-counterparts}.
The $gri$ information here for EMU-PS J213409.5--533631, EMU-PS J220026.3--561030 and EMU-PS J215026.5--621006 is taken from DES, and for SWAG-X J093803.4--015247 and SWAG-X J085234.4+062801 is taken from SDSS.}
\label{TAB:PEC-counterparts}
\end{table*}

\subsubsection{SWAG-X J084927.5--045721}
\label{SEC:ORC-1}
This ORC candidate is found in the 888~MHz SWAG-X survey.
The left panels of Figure~\ref{FIG:ORCCandidates1} show radio images of $12^{\prime} \times 12^{\prime}$ size implying no sign of association with other surrounding sources. 
In the middle panel radio contours are shown overlaid on the infrared image from WISE band W1.
The right panel shows a smaller cutout with the same size that is used to train the SOM ($5^{\prime} \times 5^{\prime}$).
The source has a near circular shape with a diameter of $\sim 50^{\prime \prime}$.
The integrated 888~MHz flux density is 228 mJy.
This source is also known as PMN J0849-0457 \citep[Parkes-MIT-NRAO Surveys;][]{wright94}.

We identify five optical/infrared sources near the ORC candidate.
The left panel of Figure~\ref{FIG:ORCs2} shows the radio contours overlaid on the DESI LS DR9 composite image using $gri$ optical bands.
Near the geometrical centre of the ORC candidate, we find a bright optical/infrared source labelled as ``A", which is
WISEA~J084927.33-045732.3, and \citep[2MASS J08492733-0457315;][]{skrutskie06}.
DESI LS DR9 gives a highly uncertain photometric redshift of $z=0.02\pm0.05$. 
The Gaia parallax ($1.9\pm0.3$ mas) and proper motion ($13.31\pm0.36$ mas/year) measurements suggest that it is a nearby Galactic star \citep[][]{brott05}.
A galaxy labelled as ``B" is located towards the south-east of ``A".
This galaxy is WISEA~J084927.80-045741.1 and also \citep[2MASX J08492779-0457412;][]{jarrett00}, with
a photometric redshift $z_{\rm ph} = 0.08\pm0.01$. 

Two more galaxies labelled as ``C" and ``D" are located at the north-east edge at photometric redshifts of $0.08\pm0.02$ and $0.08\pm0.01$, respectively. 
Galaxy ``C" is WISEA~J084928.60-045715.0 or also 2MASX~J08492860-0457152, with $z_{\rm spec} = 0.07697$ \citep[][]{jones09}.
Galaxy ``D" is WISEA~J084926.56-045715.9 or also 2MASS~J08492840-0457017.
One more galaxy labelled as ``E" (WISEA J084926.56-045715.9) is located at the north-west edge with $z_{\rm ph}=0.09\pm0.01$.
The redshifts of these four galaxies are consistent with 0.08 which may also be the redshift of the ORC candidate.
Note that the detected radio emission of this source resembles that of previously known ORCs. 
However, two collimated jets from galaxy ``B", seen in the Very Large Array Sky Survey (VLASS) 2-4 GHz images\footnote{http://cutouts.cirada.ca/} \citep[][]{lacy20}, 
suggest that it may also be a bent-tail radio galaxy with its too far outer tails forming a rare ring-like shape.
A dedicated study of this radio source may help us to understand the physics of the previously known ORC J2058--5736 that 
also have ring-shaped radio lobes \citep[][]{norris21b}.

We find 4 four galaxy clusters within the $10^{\prime}$ radius (closest one at a separation of $\sim 3^{\prime}$) of this radio source in the Canada France Hawaii Telescope Legacy Survey (CFHTLS) galaxy cluster catalogue \citep[][]{durret11}.
However they are all located at much higher redshifts, between 0.75 and 1.
We also look for possible associations with galaxy clusters in DESI survey \citep[][]{zou21} and do not find any below $z=0.5$.
However, the cluster catalogued as WHY~J084927.8--045741 with $z=0.0935$ \citep[][]{wen18} lies within the ASKAP detected emission,
and includes the group of galaxies seen in the left panel of Fig.~\ref{FIG:ORCs2}
In Section~\ref{SEC:Discussion}, we discuss a galaxy overdensity around this radio source.

\subsubsection{EMU-PS J222339.5--483449}
\label{SEC:ORC-2}
This ORC candidate is in the EMU-PS survey field and was also discovered serendipitously \citep[][]{norris22prep}.
We independently rediscover this source using our machine learning technique.
It has a near circular morphology with diameter of $\sim 80^{\prime \prime}$.
From left to right, the top panels of Figure~\ref{FIG:ORCCandidates1} show radio continuum image, radio contours overlaid on WISE-W1 infrared image, and a smaller cutout with the same size that is used to train the SOM.
The $12^{\prime} \times 12^{\prime}$ radio continuum image shows that it has no association with any of the extended radio structures in its vicinity.

We identify four optical/infrared sources near this ORC candidate.
The right panel of Figure~\ref{FIG:ORCs2} shows radio continuum contours overlaid on DES $gri$-color composite image.
Near its geometrical centre, we find an optical/infrared source labelled ``A". 
It is WISEA J222339.73-483457.9, and DESI LS DR9 gives $z_{\rm ph}=0.34\pm0.04$ for it. 
Its morphological type is not known but the colors indicate that it is a passive galaxy.

Towards the north-east edge, we find a galaxy (labelled ``B") which is WISEA J222339.53-483524.8, or also 2MASS J22233951-4835247 at $z_{\rm ph}=0.22\pm0.02$.
Near the southern edge, we identify another galaxy (labelled ``C") which is WISEA J222343.07-483440.6, or also 2MASS J22234313-4834406, at $z_{\rm ph}=0.23\pm0.01$.
Another optical/infrared source labelled as ``D" (WISEA J222337.80-483442.4) is seen due west of the radio source centre, with $z_{\rm ph}=0.33\pm0.04$.

We find one galaxy cluster at a separation of $\sim 8^{\prime}$ using the galaxy cluster catalogue from South Pole Telescope \citep[SPT;][]{bleem15}.
This cluster is both far away from ORC candidate and is located at a much higher redshift of 0.65.
We also look for possible associations with galaxy clusters in the DESI survey \citep[][]{zou21} and find one galaxy cluster at a separation of $\sim 4^{\prime}$ and $z=0.51\pm0.02$.
As the maximum redshift among all optical/infrared sources is much smaller, this galaxy cluster is not likely to be associated with the ORC candidate.
In Section~\ref{SEC:Discussion} we discuss other possibilities of association.

Other than the ORC candidates, we also find several other peculiar radio morphologies among the 0.5\% of sources with highest Euclidean distance.
Table~\ref{TAB:PEC-counterparts} shows the properties of infrared and optical sources near them.
We briefly describe these radio sources in the following sections.

\subsubsection{EMU-PS J213409.5--533631}
\label{SEC:PEC-1}
This peculiar radio source found in the EMU-PS consists of a group of distorted radio components, collectively known as 
PKS~2130--538 \citep[][]{otrupcek91}, and nicknamed ''the dancing ghosts" \citep[see Figure~21 in][]{norris21}.
This radio source has the highest Euclidean distance which means that our algorithm classifies it as the most peculiar source in EMU-PS.
The top panels of Figure~\ref{FIG:unusual_radio_shapes-1} show radio and infrared images.
The top left panel of Figure~\ref{FIG:unusual_radio_shapes-1-3big} shows radio continuum contours overlaid on the DES 3-color ($gri$) composite image ($12^{\prime} \times 7^{\prime}$).
These ''dancing ghosts" are in galaxy cluster ABELL 3785 \citep[][]{abell89}.
The twisted shape of this structure is possibly due to an interaction of a intergalactic wind with radio jets from two super massive black holes in lenticular galaxies ``A" and ``C" \citep[][]{norris21}. 
The two galaxies ``A" and ``C" shown in Figure~\ref{FIG:unusual_radio_shapes-1-3big} have reported $z_{\rm spec} =0.0763$ and $0.07836$, respectively \citep[][]{lauer14}.
The galaxy ``B" has $z_{\rm ph} = 0.07444$ \citep[][]{bilicki14}.

\subsubsection{EMU-PS J220026.3--561030}
\label{SEC:PEC-2}
This peculiar radio source also has a high Euclidean distance in the EMU-PS.
The middle panels of Figure~\ref{FIG:unusual_radio_shapes-1} show radio and infrared images.
These images imply a circular morphology where radio jets are emitted from the galaxy nucleus and may have caused the jets to have bent in nearly half circles (analogous to a rotating garden sprinkler).

We identify two galaxies near this radio source.
Near the geometrical centre of the structure, we find a bright elliptical galaxy 2MASX~J22002408-5610413 (WISEA J220024.11-561041.7) labelled as ``A" with $z_{\rm spec}=0.0757$ \citep[][]{jones09}.
Another galaxy located towards the east, labelled ``B", is 2MASX J22003234--5610273 (WISEA J220032.19-561026.0),
with $z_{\rm ph} = 0.08\pm 0.01$.
The rich galaxy cluster \citep[ABELL 3826][]{abell89} at $z=0.075$ is centred $4.2^{\prime}$ (or 0.36 Mpc) north-west of the elliptical galaxy ``A".
This suggests that the shape of this radio source is induced by the cluster environment. 
Future work should study the environmental effects leading to this shape in more detail.

\subsubsection{EMU-PS J215026.5--621006}
\label{SEC:PEC-3}
This peculiar radio source has a radio core and an extended emission towards west and north-east.
The bottom panels of Figure~\ref{FIG:unusual_radio_shapes-1} show radio and infrared images.
The middle left panel of Figure~\ref{FIG:unusual_radio_shapes-1-3big} shows radio continuum contours overlaid on the DES 3-color ($gri$) composite image ($12^{\prime} \times 12^{\prime}$).

We identify three galaxies near the radio source.
The galaxy labelled as ``A" in the center of circular structure is 2MASX J21502732-6210129 (WISEA J215027.29-621013.3)
at $z_{\rm ph}=0.07\pm0.01$.
There is a lenticular galaxy (``B") in the south-west direction where the extended emission towards the west starts and whose jet passes over the circular emission towards north-east. 
This galaxy is 2MASX J21501790-6211070 (WISEA J215017.86-621106.4) with $z_{\rm ph}=0.06\pm0.01$.
One more galaxy labelled as ``C" (2MASX J21502489-6208550, WISEA J215024.94-620854.5) has $z_{\rm ph}=0.08\pm0.01$ and is located towards the north edge of the radio source. 
However, this galaxy is unlikely to act as a host of any parts of the radio emission due to its position.
We find a previously identified galaxy group $\sim 2^{\prime}$ north-east from ``A" \citep[DZ2015 028;][]{diaz15}.
This suggests that the emission around the central galaxy is possibly due to emission from the group of galaxies.
Future work should study the group environmental effects leading to this radio shape in more detail.

\begin{figure*}[!ht]
\centering
\includegraphics[width=20cm, scale=0.5]{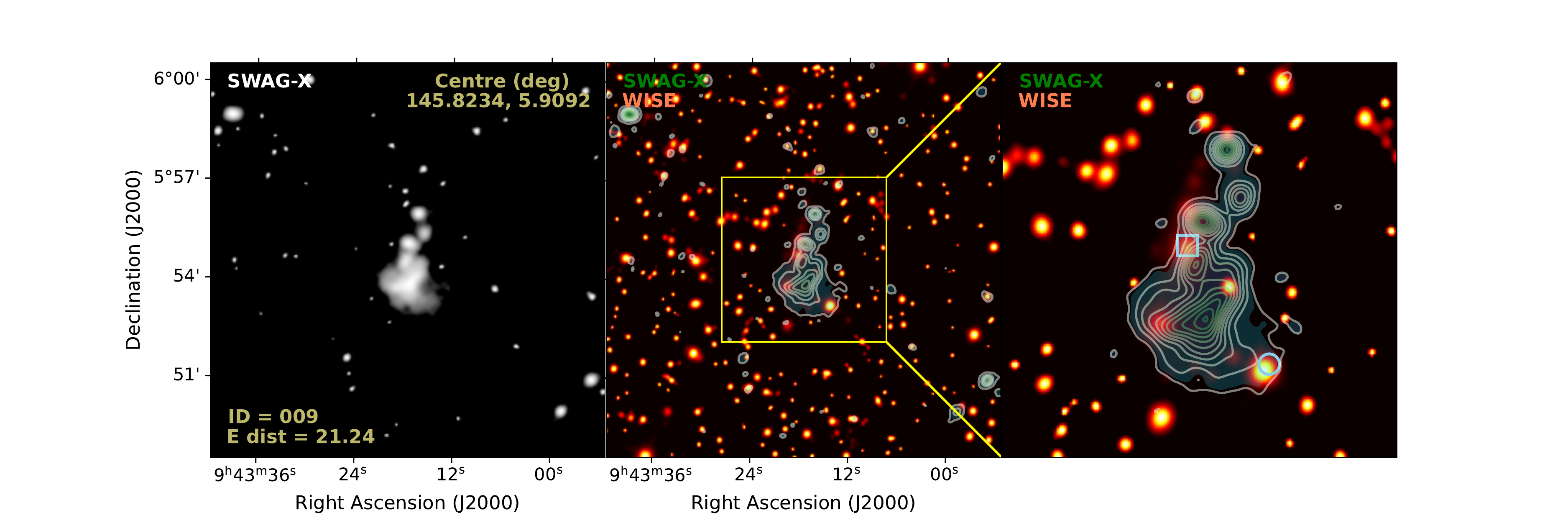}
\includegraphics[width=20cm, scale=0.5]{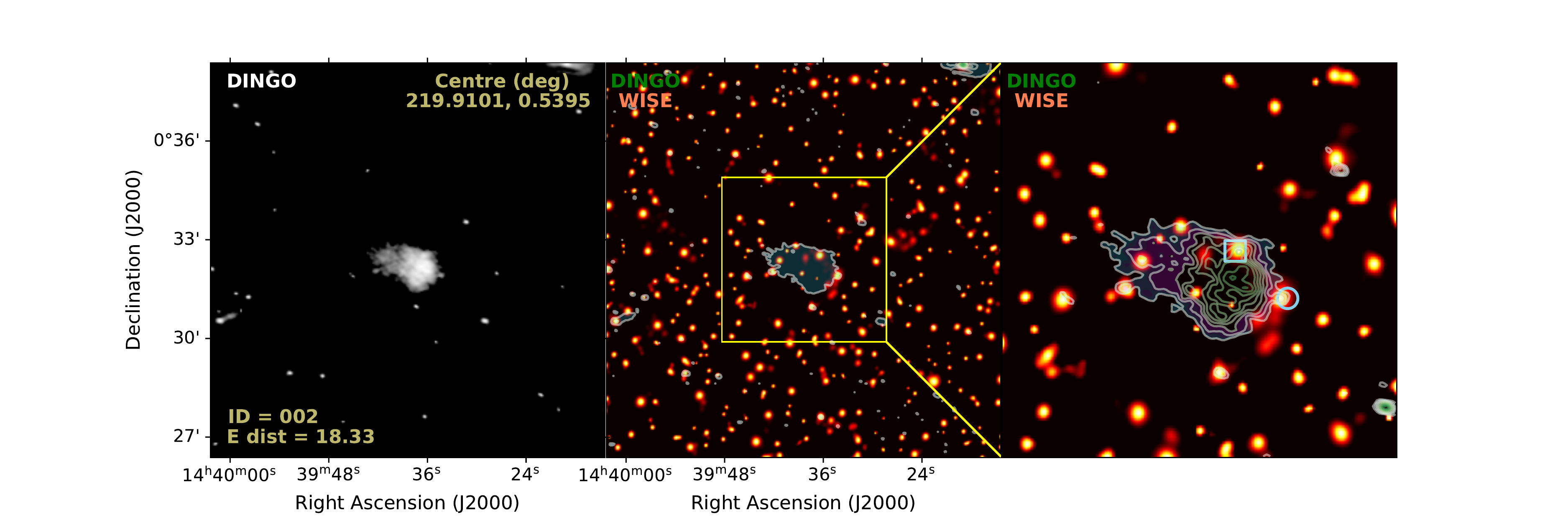}
\caption{Diffuse radio emission from galaxy clusters: The two sources are in SWAG-X (top panels) and DINGO (bottom panels) surveys (see Section~\ref{SEC:galaxyclusters}).
The description of the panels is same as Figure~\ref{FIG:ORCCandidates0}.
The sky blue square and circle in the top right panel show central BCG positions of galaxy clusters MaxBCG J145.82575+05.91142 and WHL J094322.3+055537, respectively.
In the bottom right panel, sky blue square and circle show BCG positions of galaxy clusters HSCS J143930+003220 and WHL J143934.3+003153, respectively.
Both left and middle panels are $12^{\prime} \times 12^{\prime}$ large and right panels are $5^{\prime} \times 5^{\prime}$ large which is the same size that is used to train the SOM.} 
\label{FIG:Gcl}
\end{figure*}

\subsubsection{SWAG-X J093803.4--015247}
\label{SEC:PEC-4}
This peculiar radio source is in the SWAG-X field.
The top panels of Figure~\ref{FIG:unusual_radio_shapes-2} show the radio continuum image (left), radio contours overlaid on WISE-W1 infrared image (middle), 
and a smaller cutout with the same size that is used to train the SOM (right).
The $12^{\prime} \times 12^{\prime}$ radio continuum image shows that it has no association with any of the nearby extended radio sources.

The middle right panel of Figure~\ref{FIG:unusual_radio_shapes-1-3big} shows radio continuum contours overlaid on an SDSS 3-color ($gri$) composite image.
We find an optical/infrared object labelled as ``A"(2MASS J09380334-0152480, WISEA J093803.35-015247.9) with $z_{\rm ph}=0.22\pm0.01$ near the geometrical center of the source.
This radio structure with a bright source at its center is possibly an end-on remnant radio galaxy, though it shows indications of a partial outer ring in radio emission similar to ORCs.
Future work should study this morphology in more detail.

\begin{figure*}[!ht]
\centering
\includegraphics[width=20cm, scale=0.5]{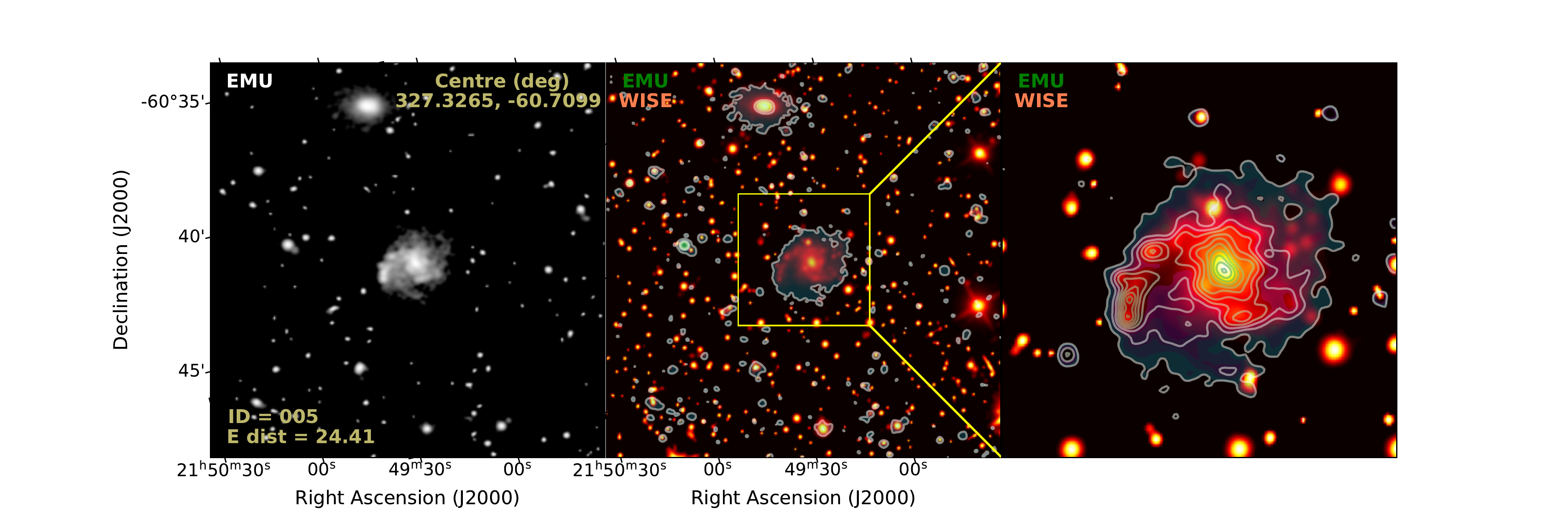}
\includegraphics[width=20cm, scale=0.5]{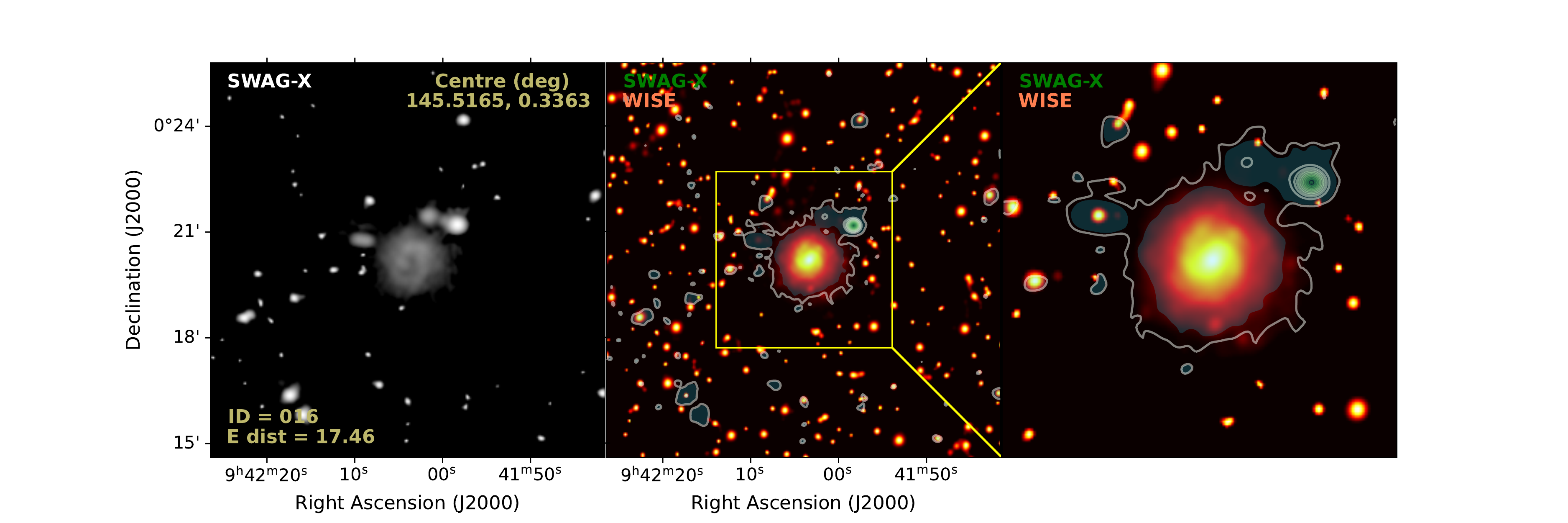}
\caption{Resolved star forming galaxies in EMU-PS survey: Top panels show NGC 7125, a spiral galaxy located at $z=0.01$. 
Bottom panels show NGC 2967, a face-on star forming spiral galaxy at $z=0.0063$.
The description of the panels is same as Figure~\ref{FIG:ORCCandidates0}.
Both left and middle panels are $12^{\prime} \times 12^{\prime}$ large and right panels are $5^{\prime} \times 5^{\prime}$ large which is the same size that is used to train the SOM.} 
\label{FIG:Resolved_galaxies}
\end{figure*}

\begin{figure*}[!ht]
\centering
\includegraphics[width=20cm, scale=0.5]{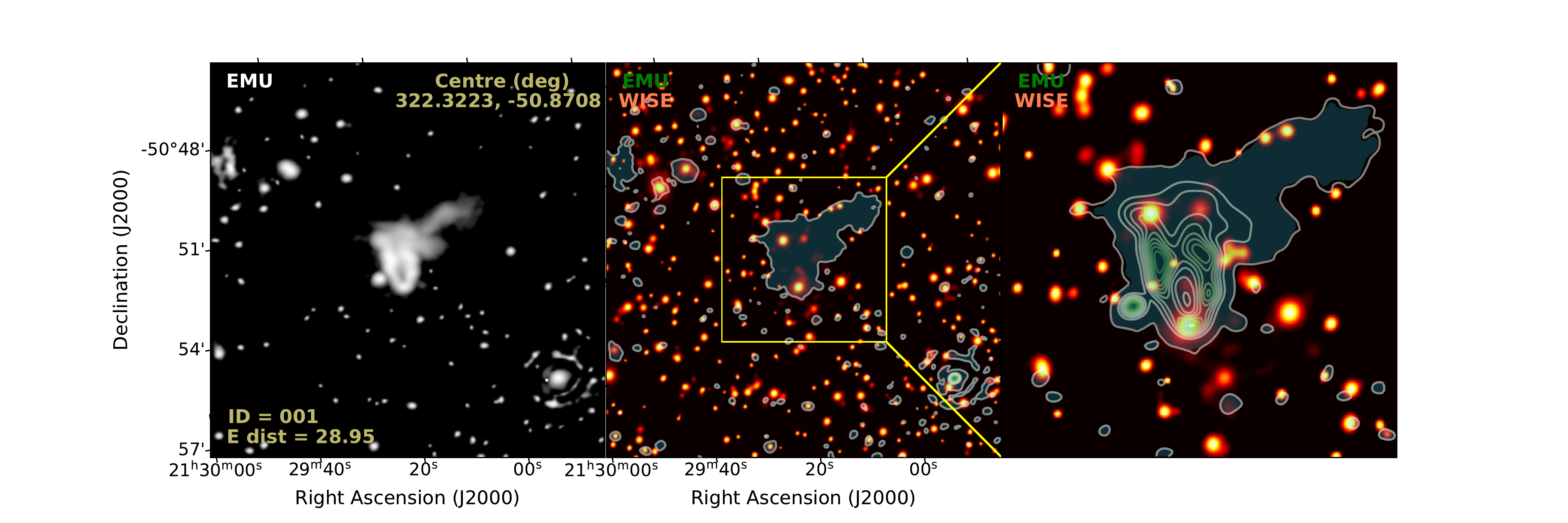}
\includegraphics[width=20cm, scale=0.5]{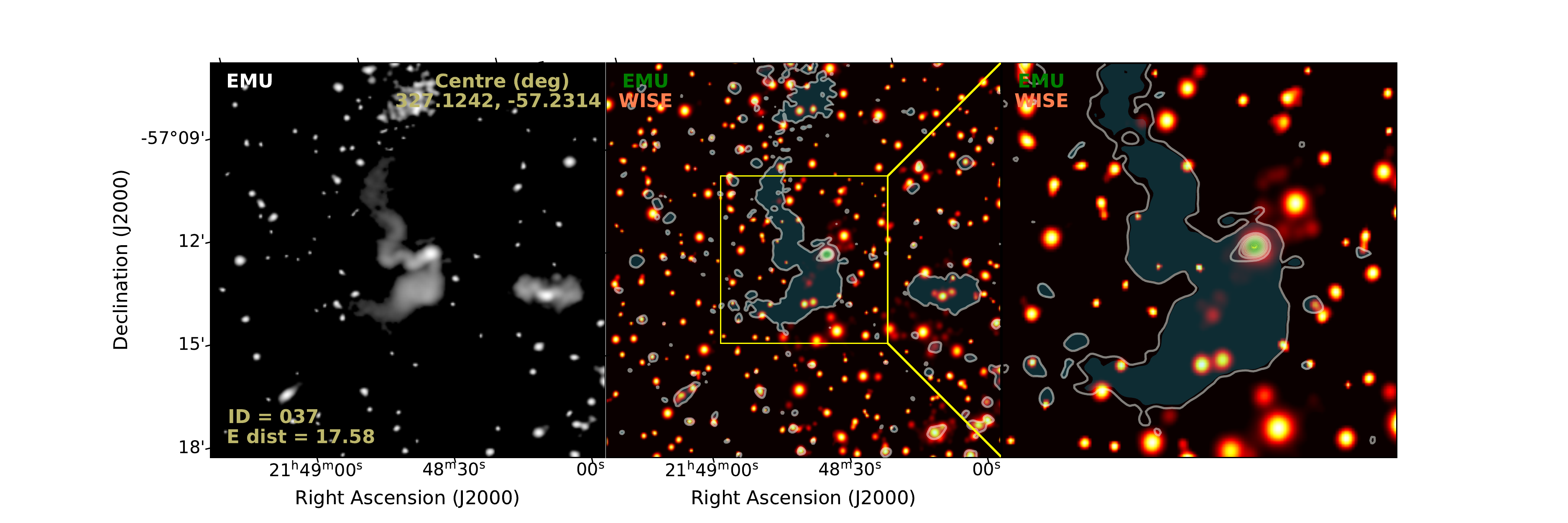}
\caption{Bent-tail (BT) radio galaxies in EMU-PS survey: Top and bottom panels show BT galaxies at $z=0.079$ and $z=0.081$, respectively.
The description of the panels is same as Figure~\ref{FIG:ORCCandidates0}.
Both left and middle panels are $12^{\prime} \times 12^{\prime}$ large and right panels are $5^{\prime} \times 5^{\prime}$ large which is the same size that is used to train the SOM.} 
\label{FIG:WATs}
\end{figure*}

\subsubsection{SWAG-X J085234.4+062801}
\label{SEC:PEC-5}
This peculiar radio morphology is also in the SWAG-X field.
From left to right, the bottom panels of Figure~\ref{FIG:unusual_radio_shapes-2} show the radio continuum image, radio contours overlaid on WISE-W1 infrared image, and a smaller cutout with the same size that is used to train the SOM.

We identify three galaxies near the edges of this structure.
The bottom panel of Figure~\ref{FIG:unusual_radio_shapes-1-3big} shows radio continuum contours overlaid on a SDSS 3-color ($gri$) composite image.
Towards the north edge, we find a galaxy ``A" (2MASX J08523573+0628209, WISEA J085235.74+062821.1) at $z_{\rm spec} = 0.15958$ (from SDSS).
Near the south-west edge, we identify a galaxy ``B" (SDSS J085232.56+062737.6, WISEA J085232.90+062731.9) at $z_{\rm ph}=0.18\pm0.06$.
Another optical/infrared object ``C" lies due south-east (2MASS J08523531+0627206, WISEA J085235.31+062720.2) and has $z_{\rm ph}=0.26\pm0.05$.  
The Gaia parallax ($3.5\pm 0.1$ mas) and proper motion ($42.9\pm 0.1$ mas/year) measurements suggests it to be a star.

The radio emission appears to be dominated by the two overlapping bright galaxies ``A" and ``B”.
In fact, galaxy ``A” with mostly compact radio emission appears to be hosting a bent-tail jet that points toward ``B” making a half circle.
The circular diffuse emission is possibly associated with ``A” and/or ``B”.
Two arcminutes north of galaxy ``A", there is an extended radio source which appears to be unrelated to the diffuse emission from this source.

\subsection{Conventional Radio Morphologies}
\label{SEC:familiar}
The ORC candidates and other peculiar radio sources discussed in the previous section are the most unusual radio morphologies in the three ASKAP pilot surveys. 
The rest of the top 0.5\% radio sources have standard morphologies with known mechanisms of formation.
These conventional sources include the diffuse emission from galaxy clusters, resolved star forming galaxies, bent-tailed galaxies and Fanaroff-Riley sources.
These sources generally have more complex shapes and larger extent compared to the typical radio galaxies, and therefore have higher Euclidean distances than the rest of the data.
The discussion of all of these sources is out of the scope of the present work.
However, we present some representative examples of these sources in this section.

\begin{figure*}[!ht]
\centering
\includegraphics[width=20cm, scale=0.5]{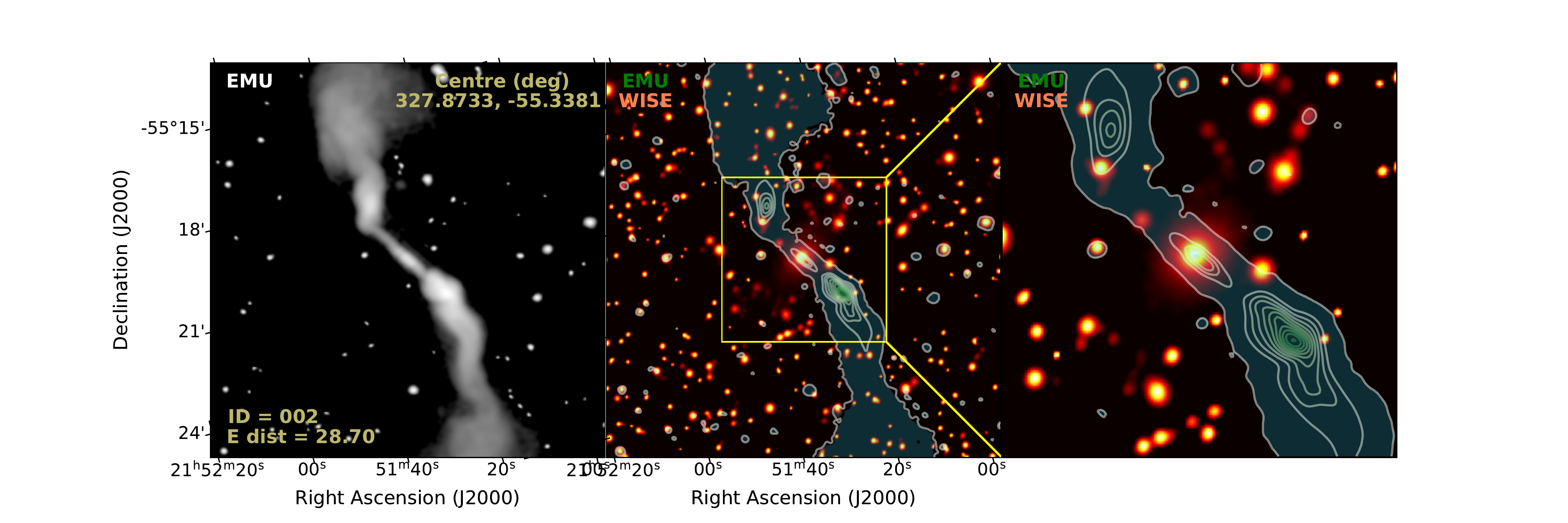}
\includegraphics[width=20cm, scale=0.5]{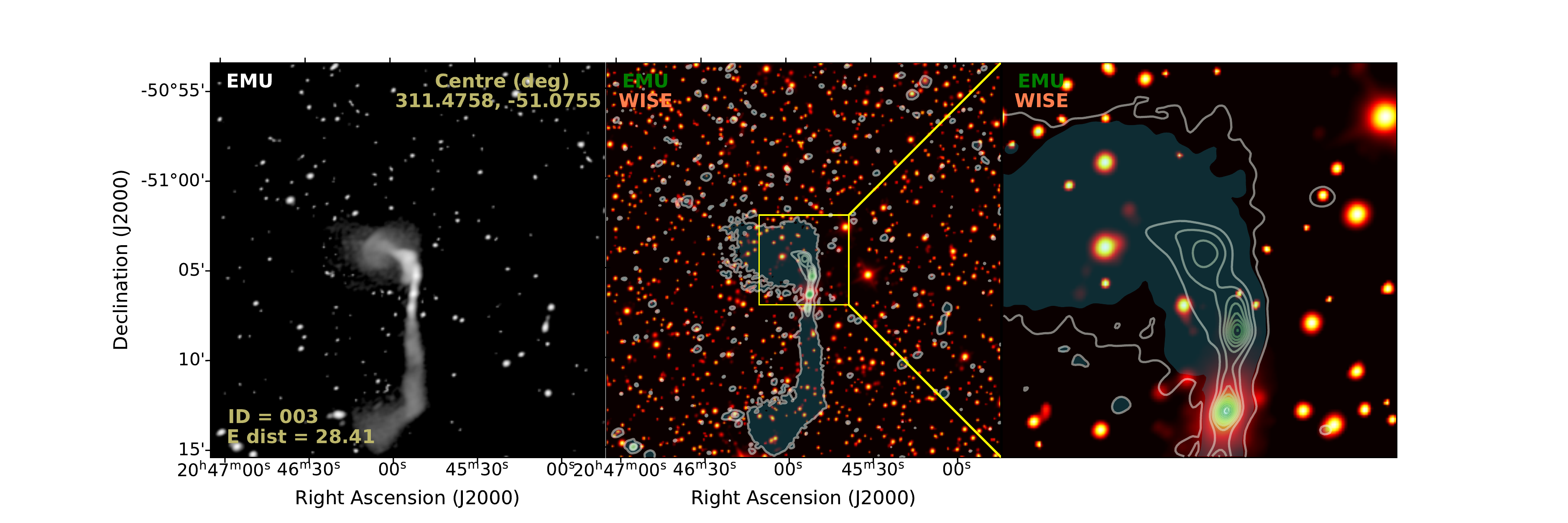}
\caption{FR-I radio galaxies in EMU-PS: the top and bottom panels show bright extended radio sources with host galaxies 2MASX J21512991-5520124 at $z=0.0388$ and 
2MASX J20455226-5106267 at $z=0.0485$, respectively.
The description of the panels is the same as for Figure~\ref{FIG:ORCCandidates0}.
The top left panel is $12^{\prime} \times 12^{\prime}$, and the bottom left panel is $25^{\prime} \times 25^{\prime}$ large. 
The right panels are $5^{\prime} \times 5^{\prime}$ large which is the same size that is used to train the SOM.} 
\label{FIG:FRI}
\end{figure*}

\begin{figure*}[!ht]
\centering
\includegraphics[width=20cm, scale=0.5]{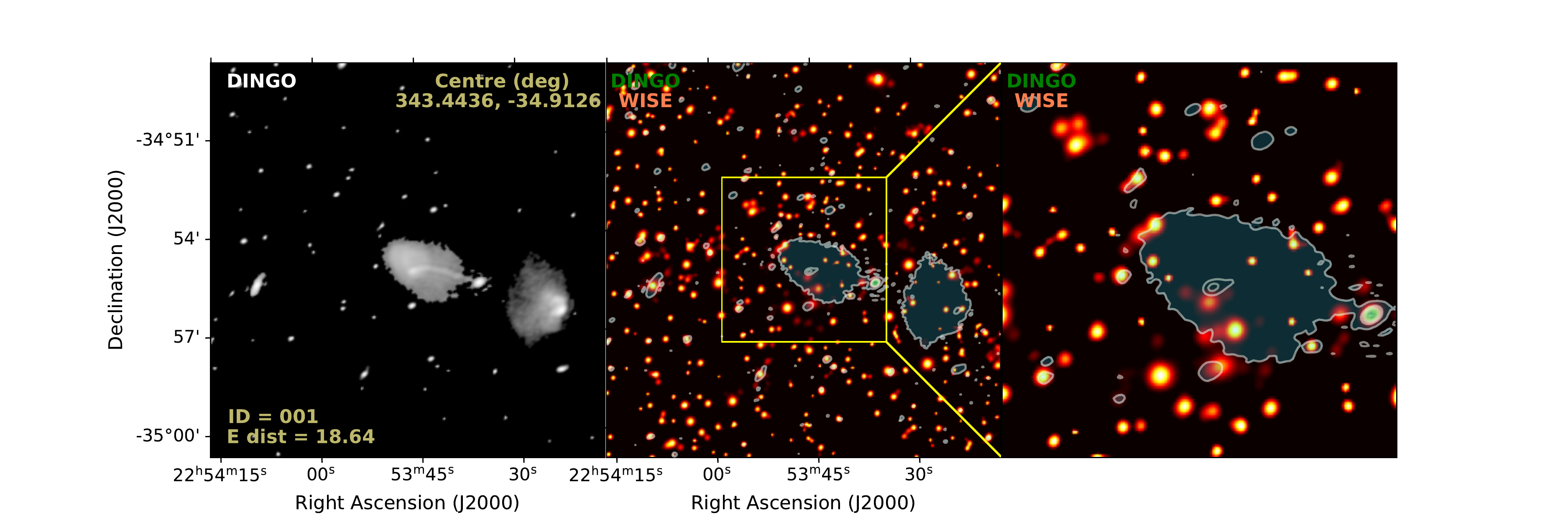}
\includegraphics[width=20cm, scale=0.5]{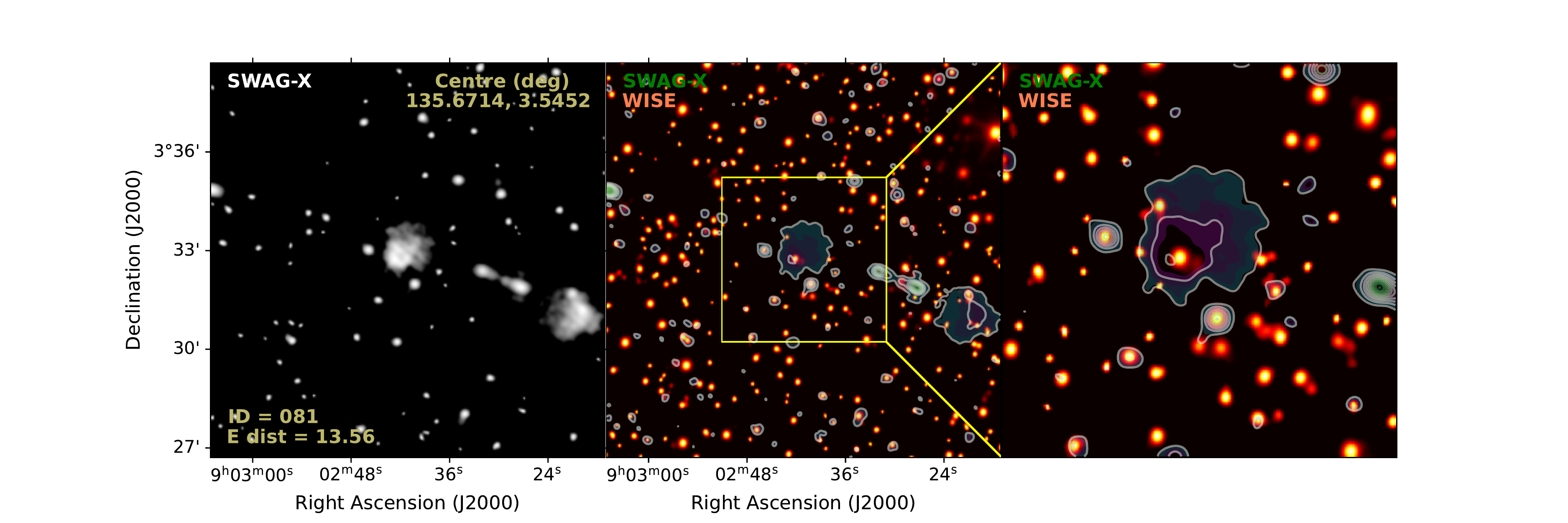}
\includegraphics[width=20cm, scale=0.5]{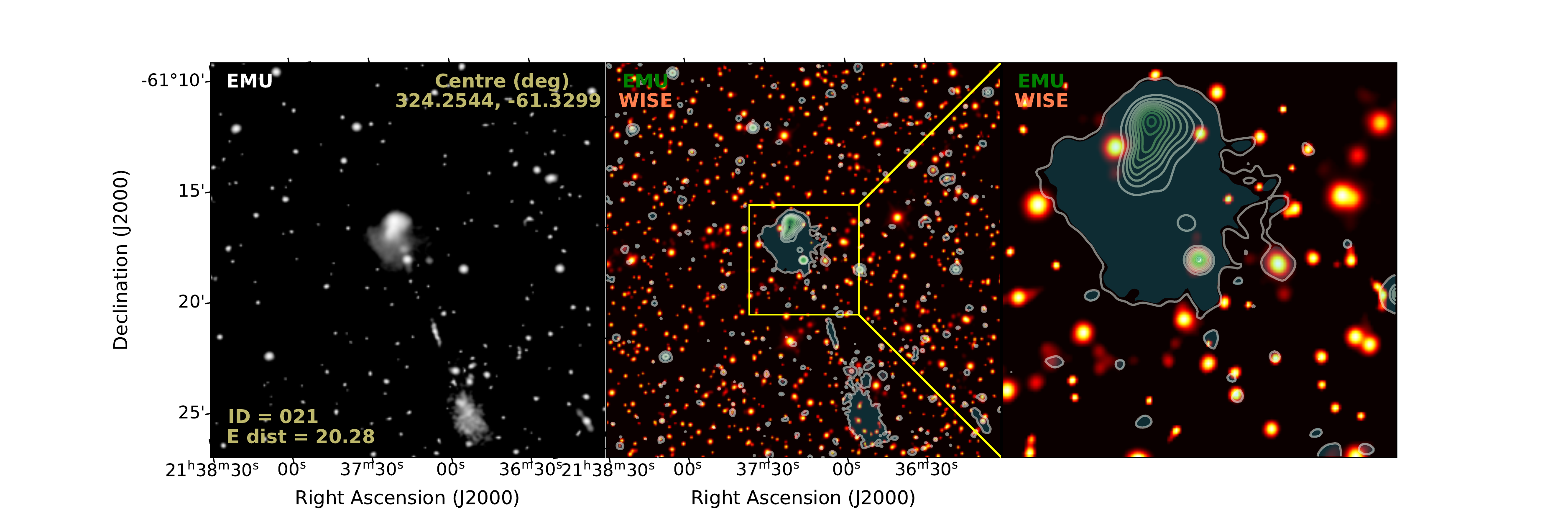}
\caption{FR-II radio galaxies in DINGO, SWAG-X and EMU-PS surveys: Top, middle and bottom panels show GRGs with host galaxies 2MASS J22533602-3455305 at $z_{\rm sp}=0.2115$, 2MASS J09022915+0332041 at $z=0.25$ and 2MASX J21365159-6125128 at $z_{\rm sp}=0.1249$, respectively.
The description of the panels is same as Figure~\ref{FIG:ORCCandidates0}.
Bottom left panels are $18^{\prime} \times 18^{\prime}$, and others are $12^{\prime} \times 12^{\prime}$ large. 
Right panels are $5^{\prime} \times 5^{\prime}$ large which is the same size that is used to train the SOM.} 
\label{FIG:FRII}
\end{figure*}

\subsubsection{Diffuse emission from galaxy clusters}
\label{SEC:galaxyclusters}
Galaxy clusters are usually detected in microwave \citep[e.g.][]{planck13-29, bleem19, hilton21}, X-ray \citep[e.g.][]{piffaretti11, liu21} and optical \citep[e.g.][]{rykoff16} wavelengths.
Galaxy clusters are known to have an overdensity of radio sources as compared to the field \citep[e.g.][]{coble07, gupta17a, gupta20b}.
Recently, a growing number of galaxy clusters are found to have sources with diffuse radio emission.
These sources are classified as radio halos, radio shocks (relics), and revived AGN fossil plasma sources \citep[e.g.][]{weeren19, giovannini20}. 
With the higher sensitivity of the new generation of radio telescopes like ASKAP, we expect to see diffuse emission from galaxy clusters.
In Figure~\ref{FIG:Gcl}, we show two such systems at very high Euclidean distances from the SWAG-X and DINGO surveys.

The top panels show diffuse emission from the galaxy cluster MaxBCG J145.82575+05.91142 identified in the SDSS survey using the maxBCG red-sequence method \citep[][]{koester07a}.
The sky-blue square in the right panel of the figure shows its brightest cluster galaxy (BCG).
This cluster is located at $z=0.094$ \citep[][]{rozo15}. 
Less than $2^{\prime}$ north-east of this system, there is an another known galaxy cluster located at $z=0.334$ \citep[WHL J094322.3+055537;][]{wen12, wen15}.
The sky-blue circle in the right panel of the figure shows its brightest cluster galaxy (BCG).

The bottom panels show a rare diffuse radio emission possibly from two galaxy clusters at different redshifts.
The radio emission has the highest Euclidean distance score which means that it is the most peculiar source in the DINGO survey.
We find galaxy clusters HSCS J143936+003231 \citep[$z=0.108$;][]{oguri18} and WHL J143934.3+003153 \citep[$z=0.15$;][]{wen15} towards the north-west and west edges of the diffuse emission.
The sky-blue cross and circle in the right panel of the figure show BCGs of HSCS J143936+003231 and WHL J143934.3+003153, respectively.
It is not clear whether both or only one of these clusters have diffuse emission towards the east of their central BCG positions.
Future dedicated work should study the radio emission these galaxy clusters in more detail.

\subsubsection{Resolved star forming galaxies}
\label{SEC:resolvedGl}
Nearby edge-on and face-on star forming galaxies are usually detected in radio continuum images and H$\alpha$ emission lines \citep[e.g.][]{pogge93, colbert96}.
In all cases, infrared and radio continuum emissions are known to be correlated \citep[e.g.][]{murphy06, vlahakis07, garn09, lacki10}.
The radio emission associated with these resolved galaxies has two well known components that correlate with the star formation rate i.e. the synchrotron emission from relativistic electrons accelerated by supernova remnants and the free–free emission emerging directly from H-II regions containing massive ionizing stars \citep[e.g.][]{condon92, murphy11, kennicutt12}.

Among the 0.5\% sources at high Euclidean distances, we find many edge-on and face-on star forming galaxies.
Figure~\ref{FIG:Resolved_galaxies} shows two such resolved star-forming galaxies in EMU-PS survey.
The top panels show NGC~7125, a spiral galaxy located at $z_{\rm sp}=0.0105$ \citep[][]{wong06}.
This galaxy is also a part of the galaxy group PGC1 0067418 NED002 \citep[][]{kourkchi17}.
The bottom panels show NGC~2967, a face-on star forming spiral galaxy at $z_{\rm sp}=0.0063$ \citep[][]{couto06}.
The star formation properties of the inner ring are known to be independent of the ring shape of this source \citep[][]{grouchy10}.

\subsubsection{Bent-tailed sources}
\label{SEC:bent-tail}
Bent-Tailed (BT) radio sources are those where radio lobes and jets are not aligned linearly with the host galaxy.
These sources are broadly classified into Wide-Angle Tail (WAT) and Narrow-Angle Tail (NAT) radio galaxies. 
WATs are usually associated with central cluster galaxies and possess a pair of well-collimated jets with small opening angles ($\leq 60^{\circ}$).
NATs have plumes of emission which are bent to such a degree
that their whole radio structure lies on one side of the optical host galaxy.
BT radio galaxies are exclusively found in the most dense environments like galaxy clusters or groups \citep[e.g.][]{mao09}.
The peculiar morphology of BT radio galaxies is typically a result of ram pressure stripping due to the relative movement of the host galaxy through an intra-cluster or intra-group medium \citep[e.g.][]{gunn72, miley72, eilek84, sakelliou2000}.

Several BT galaxies appear at high Euclidean distances among the top 0.5\% sources.
Figure~\ref{FIG:WATs} shows two such galaxies in the EMU-PS survey.
The top panels show a BT radio galaxy near the ABELL 3771 cluster at $z=0.075$ \citep[][]{martinez14}.
The bottom panels show another BT galaxy at $z=0.081$.

\subsubsection{FR-I and FR-II sources}
\label{SEC:FRI-II}
The morphologies of extended radio emission of radio galaxies are typically classified into two broad categories: 
Fanaroff-Riley Class I (FR-I) and Class II (FR-II) sources \citep[][]{fanaroff74}.
FR-I radio galaxies generally have lower radio brightness with increasing distance from the host galaxy.
FR-II radio galaxies often have linear jets that terminate in hotspots of large radio lobes.
Thus, FR-I and FR-II radio galaxies are typically described as edge-darkened and edge-brightened Active Galactic Nuclei (AGN), respectively.

We find several large scale FR-I sources among the 0.5\% sources with largest Euclidean distances, and in
Figure~\ref{FIG:FRI} we show the two topmost such FR-I sources, both found in the EMU-PS survey.
The top panels show a bright FR-I source with a total projected angular size of $\sim 12^{\prime}$. 
The host galaxy, 2MASX J21512991-5520124 with $z_{\rm sp}=0.0388$ \citep[][]{hernan95} is located in 
the galaxy cluster MCXC J2151.3-5521 \citep[][]{piffaretti11}.
The bottom panels show another FR-I radio source with host galaxy 2MASX J20455226-5106267 located at $z_{\rm sp}=0.0485$ \citep[][]{jones09} and radio emission extending over $\sim 12^{\prime}$.
Note that the cutouts used to train the SOM are on the right panels and are too small to cover the full continuum emission of FR-I sources.
Despite that the radio emission fills these cutouts to a large extent, we still find these sources at highest Euclidean distances.

Several FR-II galaxies are also found among the top 0.5\% sources. 
Figure~\ref{FIG:FRII} shows three giant radio galaxies (GRGs) in the DINGO, SWAG-X, and EMU surveys.
All these sources appear at high Euclidean distances although the information that makes them peculiar to machine learning algorithm comes from the edge-brightened hotspots as shown in the right panels.
The top panels show a FR-II source with largest angular size (LAS) of $= 4.9^{\prime}$ and projected largest linear size (LLS) of 1~Mpc. 
The host galaxy 2MASS J22533602-3455305 is located at $z_{\rm sp}=0.2115$ \citep[][]{colless03}.
This GRG in Abell 3936 has been studied in detail by \cite{seymour20}.
It shows continuous emission towards the east and a detached lobe towards the west.
The middle panels show linear radio jets from another FR-II source with host SDSS J090229.15+033204.3 (2MASS J09022915+0332041) at $z_{\rm ph}=0.25$, LAS~$= 6.8^{\prime}$ and LLS~$=1.6$~Mpc.
This source is a restarted radio galaxy that exhibits, in addition to the outer lobes, more recent double-lobed emission near the central galaxy.
The bottom panels show another FR-II source with potential host 2MASX J21365159-6125128 at $z_{\rm sp}=0.1249$ \citep[][]{colless03}, LAS~$= 11.1^{\prime}$ and LLS~$=1.49$~Mpc. 
The other potential host is  2MASS J21370099-6119472 at $z_{\rm ph}=0.277\pm0.054$ (DESI DR9) close to the north-east lobe would lead to LLS~$=2.56$~Mpc.

\section{Environment of ORC Candidates}
\label{SEC:Discussion}
The previously known three ORCs (ORC J2102–6200, ORC J2058–573 and ORC J0102–2450, see Table~\ref{TAB:all-orcs}) either lie in a significant overdensity or have a close companion \citep[][]{norris21c, norris22}.
This suggests that the environment may be important in their formation.
For the two ORC candidates from the present work, we look for possible associations with low redshift galaxy clusters in Planck \citep[][]{planck13-29}, Dark Energy Spectroscopic Instrument \citep[DESI;][]{zou21} and Meta-catalogue of X-Ray Detected Clusters of Galaxies \citep[MCXC;][]{piffaretti11} catalogues.
We do not find any galaxy cluster candidate within $10^{\prime}$ from the centre of ORCs in the redshift range of their optical sources (see Table~\ref{TAB:ORC-counterparts}).

\begin{figure}
\centering
\includegraphics[width=7.2cm, scale=0.5]{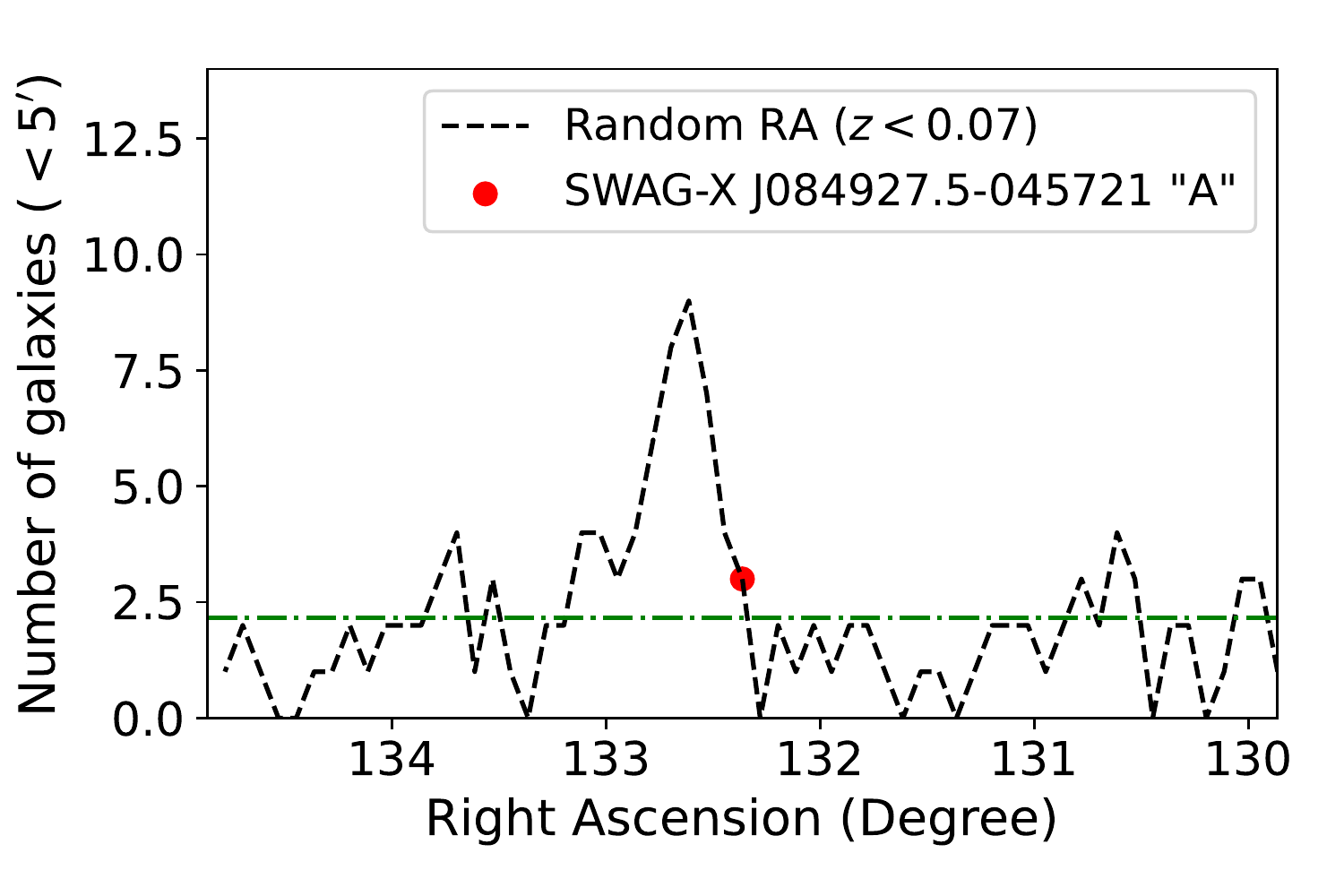}
\includegraphics[width=7.2cm, scale=0.5]{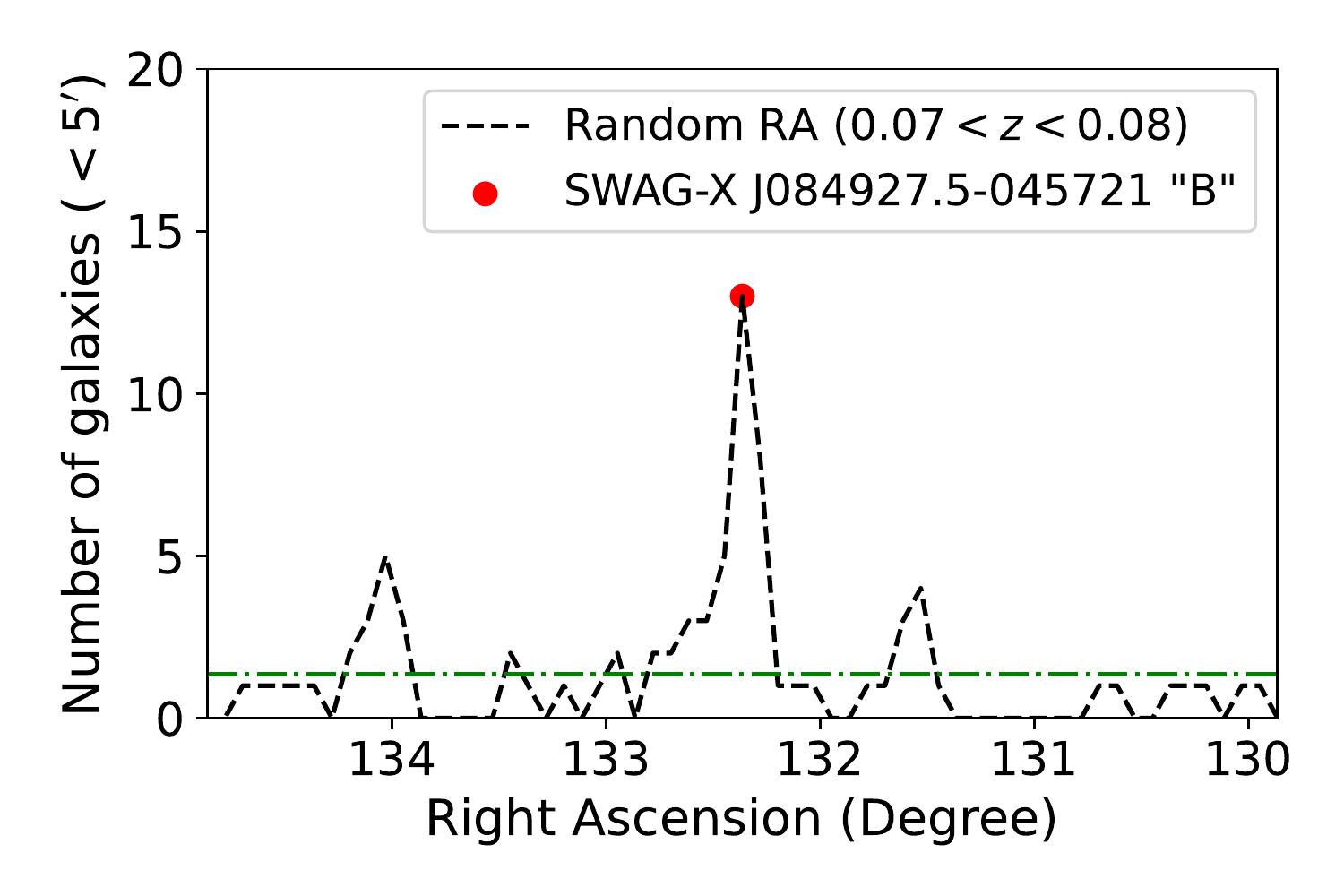}
\includegraphics[width=7.2cm, scale=0.5]{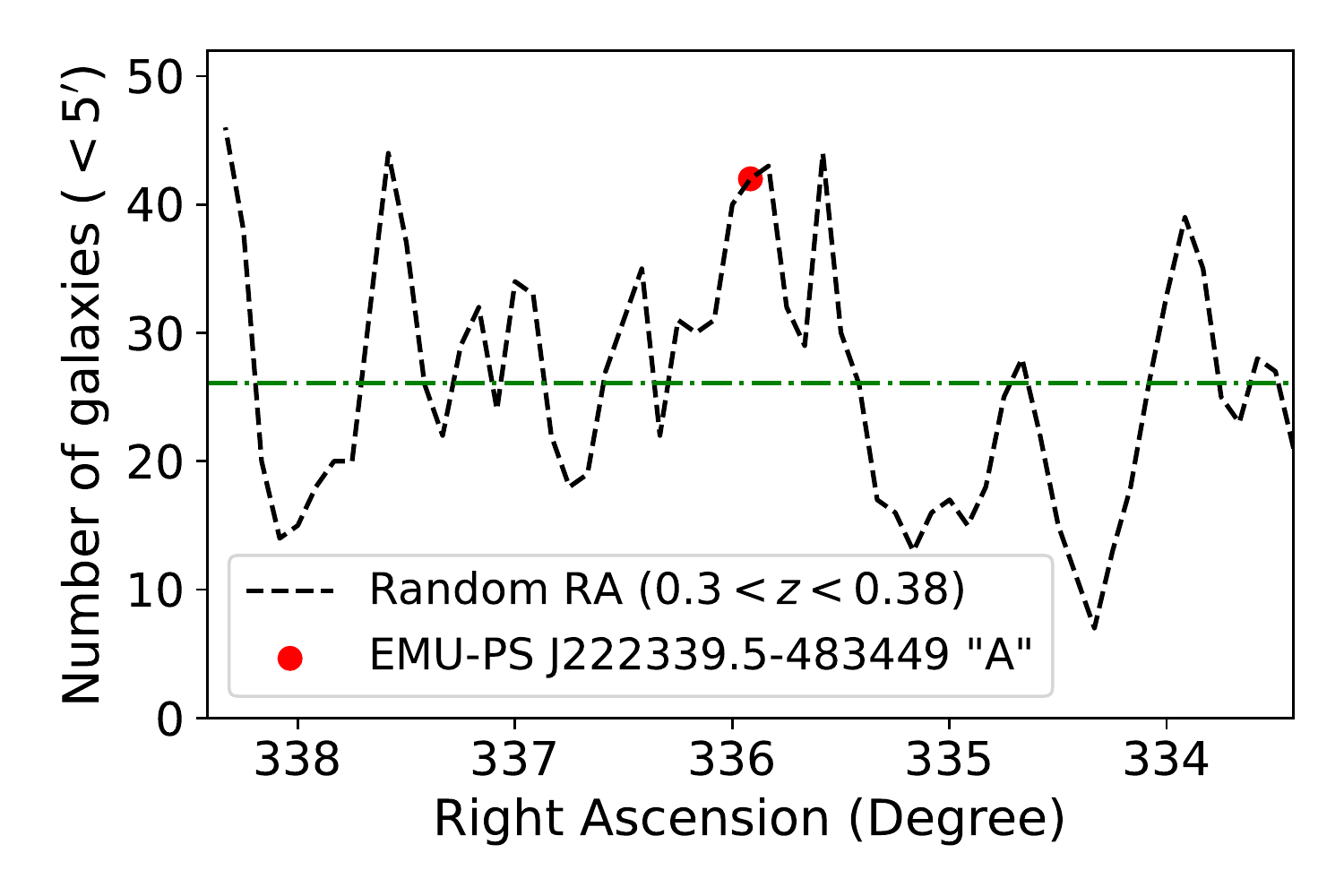}
\includegraphics[width=7.2cm, scale=0.5]{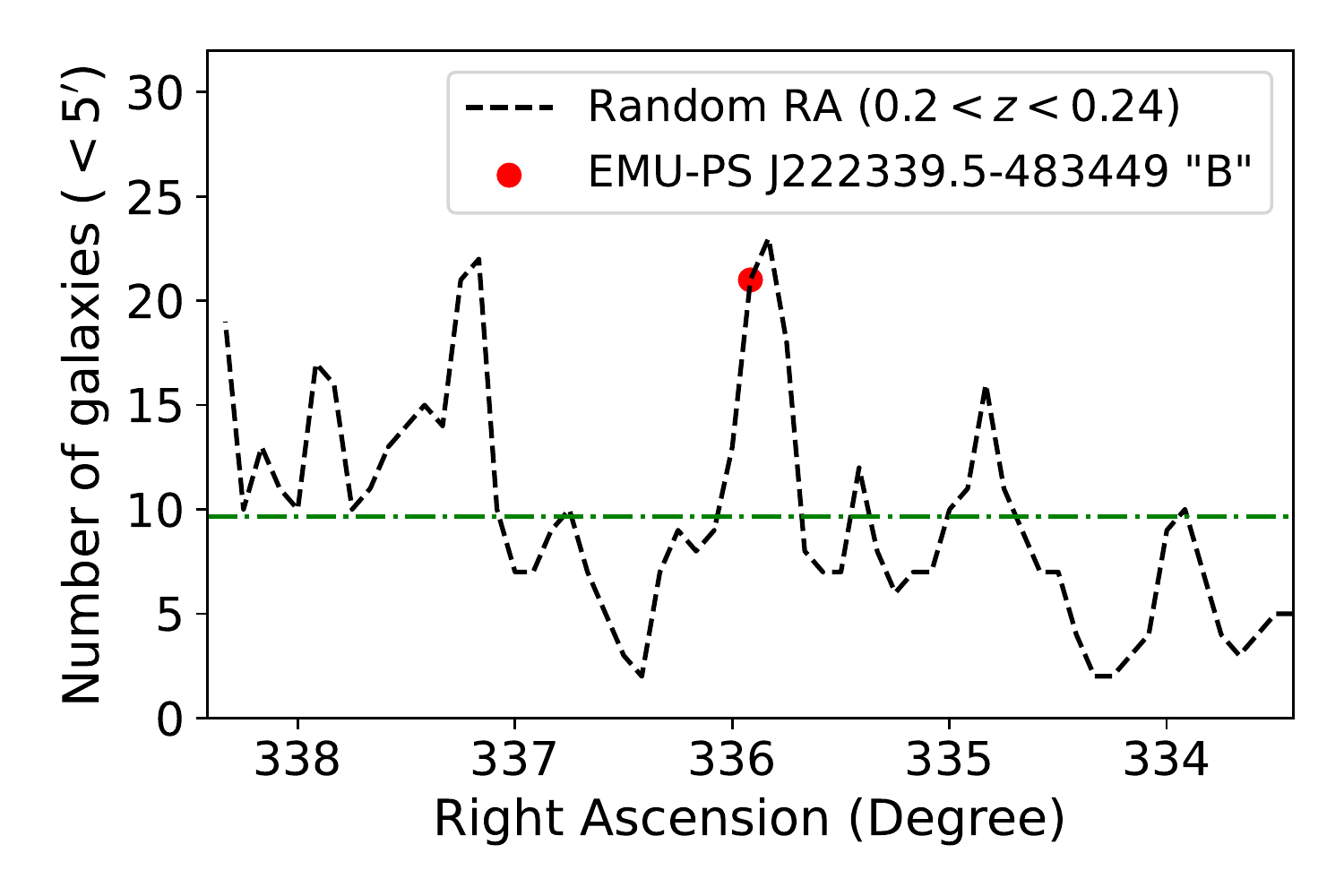}
\caption{Number of DESI DR8 galaxies in a circle of $5^{\prime}$ radius centered at the sources in ORC candidates (red circles). 
The top two panels show the galaxy number density around SWAG-X J084927.5-045721 ``A" and ``B", and the bottom two panels for EMU-PS J222339.5--483449 “A" and “B" objects.
For comparison with field densities around these sources, we show galaxy counts in circles of the same radius sliding over the RA range indicated on the X-axis but keeping the Dec fixed (black dashed lines).
The green dot-dashed lines show average number of galaxies in the RA range.
Given the redshift uncertainties of ORC candidate sources (Table~\ref{TAB:ORC-counterparts}), we restrict DESI galaxies within $z<0.07$, $0.07<z<0.09$, $0.3<z<0.38$ and $0.2<z<0.24$ (top to bottom panels).
The median redshift uncertainties of all sources in the RA and the redshift ranges are 0.0093, 0.024, 0.066, and 0.046 (top to bottom panels).
This shows an overdensity of galaxies near all objects except the SWAG-X J084927.5-04572 ``A", and suggests a possibility of unknown intergalactic physics that is shaping the circular morphologies of these ORC candidates.
}
\label{FIG:ORC1-2-overdensity}
\end{figure}

Galaxy cluster catalogues do not necessarily have group scale systems with fewer number of galaxies.
We therefore explore the overdensities of galaxies at the positions of two ORCs
using the photometric redshift catalogue from DESI DR8 \citep[][]{zou20}. 
We estimate the number of galaxies in a circle of $5^{\prime}$ radius of ``A" and ``B" galaxies near both ORC candidates (see Table~\ref{TAB:ORC-counterparts} and Fig~\ref{FIG:ORCs2}).
We chose these two galaxies as they have different redshifts that are not consistent with each other.
Redshifts of other galaxies are consistent with either of these sources.
We use their photometric redshift uncertainty as the redshift range to estimate overdensities.
We restrict the DESI photo-z catalogue within $z<0.07$ and $0.07<z<0.09$ for ``A" and ``B" galaxies of SWAG-X J084927.5--045721, respectively.
Although source ``A" here is expected to be a Galactic star (see Section~\ref{SEC:ORC-1}), nevertheless, we estimate overdensities near it due to its location at the centre of the ORC candidate.
For sources ``A" and ``B" in EMU-PS J222339.5–483449, subsets of the DESI photo-z catalogue with $0.3<z<0.38$ and $0.2<z<0.24$, respectively, were used.
We also estimate the number of galaxies in circles of radius $5^{\prime}$ sliding in a continuously increasing RA range (with $5^{\prime}$ RA increments and keeping Dec fixed) to compare the field number density with the density near ORC candidates.

The top panel of Figure~\ref{FIG:ORC1-2-overdensity} shows the number of galaxies within a circle of $5^{\prime}$ radius in the specified $z$ range.
The red circle shows that there are only 3 galaxies in $5^{\prime}$ radius from ``A" and the green dot-dashed line indicates an average galaxy count of 2.2 for the field.
Thus if the true redshift of ORC candidate is equivalent to that of ``A", the inter-galactic environment may not be the reason for its circular morphology.
The second panel from the top shows the galaxy counts for source ``B".
The red circle shows that there are 13 galaxies in $5^{\prime}$ radius and the green dot-dashed line indicates an average number of 1.3 for the field. 
This implies that the ORC candidate if located at its redshift could have formed its circular morphology under the impact of yet unknown inter-galactic processes.

The bottom panels of Figure~\ref{FIG:ORC1-2-overdensity} show the number of galaxies in a circle of $5^{\prime}$ radius of EMU-PS J222339.5--483449 ``A" and ``B" sources.
The red circles show that there are 42 and 21 galaxies in $5^{\prime}$ radius of ``A" and ``B", respectively.
The average number of galaxies around ``A" and ``B" are 26.1 and 9.7, respectively.
Although the two sources have very different redshifts, the high number density around both implies that the ambient galaxy density may be an important aspect for the circular morphology of EMU-PS J222339.5--483449.
We note that understanding the physics of the origin of ORCs is beyond the scope of the present work that focuses on the detection of peculiar systems.
Future work should study each of these rare circular and other peculiar sources to understand the physical mechanism behind their origin.

\section{Summary}
\label{SEC:Summary}
We present a machine learning method to search for the rarest and most interesting sources in ASKAP continuum radio surveys.
We use PINK which is an implementation of self-organising maps and accounts for the affine transformations of astronomical images in an efficient manner.
We train the machine learning algorithm using $\sim 42,000$ cutouts ($5^{\prime} \times 5^{\prime}$) at the positions of complex radio sources 
(sources with more than one components) in the EMU pilot survey.
The trained model is then used to map these sources into a $10\times 10$ lattice of neurons according to the relative similarity between the sources.
We use an Euclidean distance metric to compute the similarity in an unsupervised manner.
The Euclidean distance metric is then used to identify the rarest and most interesting sources in the survey.

We select a small fraction of radio sources with highest Euclidean distances for visual inspection.
We chose the top 0.5\% complex sources that amounts to 200 sources at high Euclidean distances in the EMU-PS field.
Radio sources among this cut include previously discovered circular radio sources in the EMU-PS survey.
These circular sources are also known as Odd Radio Circles (ORCs) and were previously discovered using a dedicated visual inspection of the EMU-PS field.
In addition to EMU-PS, we also search for interesting radio sources in two other ASKAP surveys, and to this end
we map complex sources in the SWAG-X and DINGO pilot surveys to the trained lattice.
Note that the training is done only once using EMU-PS and the trained model is used to map sources from the EMU-PS as well as the other two surveys.
For the SWAG-X and DINGO pilot surveys, we inspect 100 and 20 (top 0.5\%) complex radio sources, respectively.

Among these top 0.5\% complex sources at high Euclidean distances we find two new ORC candidates, namely  SWAG-X J084927.5--045721 and EMU-PS J222339.5--483449 (see Section~\ref{SEC:ORCs}).
We identify host galaxies at the positions of these ORC candidates from literature and by using multiwavelength data from AllWISE, DES, SDSS and other serendipitous surveys.
Using the DESI DR8 galaxy catalogue, we find that both ORC candidates are possibly located at local overdensities.
Other than these ORC candidates, we present five more peculiar radio sources with rare morphologies.
Future work should study each of these peculiar sources to understand the physical mechanism behind their origin.
The rest of the the top 0.5\% complex radio sources have conventional morphologies.
In the present work, we show some representative examples of these sources which include diffuse emission from galaxy clusters, resolved star-forming galaxies, and bent-tailed, FR-I and FR-II radio galaxies (see Sections~\ref{SEC:galaxyclusters}, \ref{SEC:resolvedGl}, \ref{SEC:bent-tail} and \ref{SEC:FRI-II}).
A useful list of all sources requires additional work to cluster multiple component radio sources and identify them with optical/infrared host galaxies. The development of such a clustering method is currently in progress and we plan to discuss it in detail in our future work.
The scope and intent of the present work is to develop a method to discover unusual objects in the next generation radio surveys that are expected to detect multi-million radio sources.

Our machine learning method detects previously known ORCs and new ORC candidates among the top 0.5\% complex sources which amounts to only 200 systems for EMU-PS.
This number is quite small for the pilot ASKAP surveys investigated here.
The full surveys will produce many more possibilities of finding rare sources.
For instance, the EMU survey is expected to produce a catalogue of 40 million radio components \citep[][]{norris11} with 18\% components associated with complex sources.
A fraction of 0.5 per cent of complex sources would lead to $\sim 36,000$ sources for visual inspection.
As the latter will be highly time-consuming,
future work should further improve the machine learning method by implementing new techniques to automate the discovery of rare objects in big surveys.

Future work should also investigate the means to include small scale (few arcseconds) and large scale (several arcminutes e.g. some FR-I and FR-II) radio source images in the training sample.
As mentioned in Section~\ref{SEC:method}, only 1 in 20,000 sources have a larger extent than the $5^{\prime}$ image size used for training the machine.
However, we find these sources at high Euclidean distances where only a small part of the continuum emission is contained in the $5^{\prime}\times 5^{\prime}$ cutout.
In addition, several images at high Euclidean distances have multiple radio sources.
These images are filled with several point-like as well as small scale double lobed galaxies.
Future work should study ways to reduce the number of such images.
Finally, while here we rely on the source catalogues to find peculiar radio sources,
future work should develop ML models based only on the full survey images to localise and detect rare morphologies.

\section{Acknowledgements}
The Australian SKA Pathfinder is part of the Australia Telescope National Facility (https://ror.org/05qajvd42) which is managed by CSIRO. Operation of ASKAP is funded by the Australian Government with support from the National Collaborative Research Infrastructure Strategy. ASKAP uses the resources of the Pawsey Supercomputing Centre. Establishment of ASKAP, the Murchison Radio-astronomy Observatory and the Pawsey Supercomputing Centre are initiatives of the Australian Government, with support from the Government of Western Australia and the Science and Industry Endowment Fund. We acknowledge the Wajarri Yamatji people as the traditional owners of the Observatory site.
The photometric redshifts for the Legacy Surveys (PRLS) catalogue used in this paper was produced thanks to funding from the U.S. Department of Energy Office of Science, Office of High Energy Physics via grant DE-SC0007914.
This research has made use of the NASA/IPAC Extragalactic Database (NED), which is operated by the Jet Propulsion Laboratory, California Institute of Technology, under contract with the National Aeronautics and Space Administration.
NG acknowledges support from CSIRO’s Machine Learning and Artificial Intelligence Future Science Platform. HA has benefited from grant CIIC 138/2022 of Universidad de Guanajuato, Mexico.

\bibliography{ASKAP_PASA}

\label{lastpage}
\end{document}